\documentclass[aps,pre,preprint,groupedaddress,showpacs]{revtex4-1}
\usepackage[dvips]{graphicx}
\usepackage{textcomp}
\usepackage{amssymb}
\newcommand{\pvec}[1]{\vec{#1}\mkern2mu\vphantom{#1}}

\newcommand{\mc}{\multicolumn}

\begin{document}

\title{
\Large\bf Monte Carlo study of an improved clock  model in three
  dimensions}

\author{Martin Hasenbusch}
\email[]{M.Hasenbusch@thphys.uni-heidelberg.de}
\affiliation{
Institut f\"ur Theoretische Physik, Universit\"at Heidelberg,
Philosophenweg 19, 69120 Heidelberg, Germany}
\date{\today}

\begin{abstract}
We study a generalized clock model on the simple cubic lattice.
The parameter of the model can be tuned such that the amplitude of 
the leading correction to scaling vanishes.
In the main part of the study we simulate the model with $Z_8$
symmetry. At the transition, with increasing length scale, 
$O(2)$ symmetry emerges.
We perform Monte Carlo simulations using a hybrid of local Metropolis 
and cluster algorithms of lattices with a linear size up to $L=512$. 
The field variable requires less memory and the updates 
are faster than for a model with $O(2)$ symmetry at the microscopic level. 
Our finite size scaling analysis yields
accurate estimates for the critical exponents of the three-dimensional
XY-universality class. In particular we get $\eta=0.03810(8)$,
$\nu=0.67169(7)$, and $\omega=0.789(4)$. Furthermore we obtain estimates
for fixed point values of phenomenological couplings and critical
temperatures.
\end{abstract}

\keywords{}
\maketitle

\section{Introduction}
In the neighborhood of a second order phase transition, thermodynamic quantities 
diverge, following power laws. For example the correlation length $\xi$ behaves
as 
\begin{equation}
\label{xipower} 
\xi = a_{\pm} |t|^{-\nu}  \; \left(1+b_{\pm} |t|^{\theta} + c t + ...\right)
\;,
\end{equation}
where $t=(T-T_c)/T_c$ is the reduced temperature. The subscript $\pm$ of the
amplitudes $a_{\pm}$ and $b_{\pm}$ indicates the high ($+$) and the low
($-$) temperature phase, respectively.
Second order phase transitions are grouped into universality classes. For all 
transitions within such a class, critical exponents like $\nu$ assume the identical
value. 
These power laws are affected by corrections. There are non-analytic or confluent 
and analytic ones. The leading corrections are explicitly given in eq.~(\ref{xipower}). 
Also correction exponents such as $\theta=\omega \nu$ are universal. 
For the system discussed here, $\theta \approx 0.5$. Amplitudes such as 
$a_{\pm}$ and $b_{\pm}$ depend on the microscopic details of the system.
However certain combinations, so called amplitude ratios, assume universal
values.
Universality classes are characterized by the symmetry properties of the 
order parameter at criticality, the range of the interaction and the spacial 
dimension of the system. For reviews on critical phenomena see for
example \cite{WiKo,Fisher74,Fisher98,PeVi}.

Note that in general the symmetry properties of the order parameter 
can not be naively inferred from the microscopic properties of the system.
In particular a symmetry might emerge that is not present in the 
classical Hamiltonian. For example, in the model studied here, the symmetry 
is enhanced from $Z_N$ to $O(2)$ at the critical point.  
At the $O(2)$-invariant Wilson-Fisher fixed point in three dimensions, 
a perturbation that
breaks the $O(2)$-invariance down to $Z_N$-invariance is irrelevant in the 
sense of the renormalization group (RG) for
$N \ge 4$. See ref. \cite{Debasish} and references therein. Monte Carlo 
studies have shown that the transition of $N$-state clock models on the simple 
cubic lattice are in the domain of attraction of the $O(2)$-invariant
fixed point for $N \ge 5$. See for example ref. \cite{HoSu03}. 
The major part of our simulations are performed for $N=8$. The 
related RG-exponent takes the value $y_{N=8} =-5.278(9)$, 
see table II of ref. \cite{Debasish}.
Therefore, the deviations from $O(2)$-invariance rapidly vanish 
with increasing lattice size and can be safely ignored in the finite size
scaling analysis at the critical point.
The  $Z_N$-invariant perturbation is dangerously irrelevant.
In the low temperature phase, in the thermodynamic limit, the spontaneous 
magnetization might only assume one of the $N$ directions, that are 
preferred by the Hamiltonian. See for example ref. \cite{Sandvik07} and 
references therein. 
In the present work, we focus on the critical point and consider a model
with $Z_N$-symmetry mainly for technical reasons. Less memory is needed to 
store the configurations and the updates require less CPU-time than for a model
with $O(2)$ symmetry.

The three-dimensional XY-universality class has attracted much attention, 
since the $\lambda$-transition of $^4$He, which is well studied experimentally,
is supposed to share this universality class.
The most accurate result for the exponent $\alpha$ of the specific heat 
is obtained from an experiment under the condition of microgravity
\cite{Lipa96,Lipa00,Lipa03}:
\begin{equation}
\alpha = - 0.0127(3)  \;\;,
\end{equation}
which corresponds to $\nu = (2-\alpha)/d = 0.6709(1)$. 

The three-dimensional XY-universality class has been studied by using various 
theoretical approaches. 
For example field theoretic methods, high and low temperature series expansions 
and Monte Carlo simulations of lattice models. 
A few representative results for critical exponents are given in table \ref{methods}.
Note that other critical exponents can be obtained from $\nu$ and $\eta$
by using scaling relations.
For more comprehensive collections see table I of ref. \cite{XY1},
table 19 of ref. \cite{PeVi}, or table I of ref. \cite{Xu19}.
Recently great progress has been achieved by using the so called 
conformal bootstrap method. In particular in the case of the three-dimensional
Ising universality class, the accuracy that has been reached for critical 
exponents clearly surpasses that of other theoretical methods. See ref. 
\cite{Simmons-Duffin:2016wlq} and references therein. For the $XY$ and the 
$O(3)$ universality classes in three dimensions the results obtained so far, 
are less precise. 
The estimates given in table \ref{methods} are derived
from the numbers for the scaling dimensions $\Delta_{\phi}$ and $\Delta_s$
given in ref. \cite{Kos:2016ysd}.  Note that after we had put the first version 
of this paper on arXiv, ref. \cite{che19} has been submitted. The accurate 
results, that come with rigorous bounds, are in nice agreement with ours.
In the last row of table \ref{methods} we report as preview the results 
of the present work. We fully confirm ref. \cite{XY2}, the discrepancy with
the experiment \cite{Lipa96,Lipa00,Lipa03} remains.

\begin{table}
\caption{\sl \label{methods}
A few representative results for the critical exponents $\nu$, $\eta$ and 
$\omega$ for the universality class of the three-dimensional XY model
obtained by various theoretical methods. MC+HT means that 
Monte Carlo simulations and the analysis of high temperature expansions
have been combined to analyze the lattice models under consideration.
}
\begin{center}
\begin{tabular}{ccclll}
\hline
Ref.&method & year & \mc{1}{c}{$\nu$}&\mc{1}{c}{$\eta$}&\mc{1}{c}{$\omega$}  \\
\hline
\cite{GuZi98} & $\epsilon$-expansion & 1998 & 0.6680(35) & 0.0380(50) & 0.802(18) \\
\cite{GuZi98} & 3D-expansion         & 1998 & 0.6703(15) & 0.0354(25) & 0.789(11) \\
\cite{XY2} & MC+HT       & 2006 & 0.6717(1)  & 0.0381(2)  & 0.785(20) \\  
\cite{Xu19} &  MC    &  2019 & 0.67183(18)  & 0.03853(48) &  0.77(13)  \\
\cite{Kos:2016ysd}& conformal bootstrap & 2016 &0.6719(11) &0.03852(64) & \\
\cite{che19}     & conformal bootstrap & 2019 & 0.67175(10) & 0.038176(44)  & \\
present work     & MC         & 2019  & 0.67169(7) &0.03810(8) & 0.789(4) \\ 
\hline
\end{tabular}
\end{center}
\end{table}

An important feature of refs. \cite{XY1,XY2} is that so called 
improved models are studied. One considers models that have  
one parameter in addition to the inverse temperature and the external 
field. On the critical line, the amplitude $b_{\pm}$ of leading corrections
to scaling, eq.~(\ref{xipower}), depends on this parameter. If there
exists a value of the parameter with $b_{\pm}=0$, RG-theory predicts
that the same holds for all quantities that are singular at the transition.
In the following we shall call a model with $b_{\pm}=0$ an improved model.
The idea had been exploited first by using high temperature series 
expansions of such models \cite{ChFiNi,FiCh}. For early  Monte Carlo
simulations of improved models sharing the universality class of the 
three-dimensional Ising model see for example refs. 
\cite{Bloete,Ballesteros,KlausStefano}.  

In the present work, we study a generalization of the $N$-state clock model, 
which is closely related with the ddXY model that has been studied in 
refs. \cite{XY1,XY2}. In addition to the $N$ values on the unit circle, 
the field variable might take the value $(0,0)$ in the center of the 
circle. We refer to this model as $(N+1)$-state clock model. 
Its precise definition is given in section \ref{theModel} below.

We study the model by using finite size scaling (FSS) \cite{Barber}.
The outline of the study builds upon our previous work on critical
phenomena, see for example refs. 
\cite{KlausStefano,myPhi4,Tibor,XY1,XY2,ourdilute,Ha10}, to give only a few. 
An important feature of these studies is that in addition to the Binder
cumulant \cite{Bi81}, other dimensionless ratios like the second 
moment correlation length over the linear lattice size $\xi_{2nd}/L$
or the ratio of the partition functions for periodic and anti-periodic
boundary conditions $Z_a/Z_p$ are exploited. The comparison of results
obtained from these different quantities allows us to estimate systematical
errors that are caused by subleading corrections that are not explicitly taken 
into account in the fits. 

The purpose of the present work is twofold. First we improve the accuracy 
of the critical exponents of the three-dimensional XY universality class.
These results provide a benchmark for future theoretical progress 
achieved by the conformal bootstrap or other methods. In fact, in the case 
of ref. \cite{che19} this already occurred.
Second we provide 
non-universal results, like for example inverse critical temperatures,
which are important groundwork for future studies. In particular 
we intend to compute the structure constants using a similar approach as 
in ref. \cite{myStructure} for the Ising universality class. Furthermore
the improved $(N+1)$-state clock model should be a good starting point to study
the symmetry properties of the order parameter in the low temperature 
phase. 

The outline of the manuscript is the following: 
In section \ref{theModel} we define the  model and the observables that we
measured. We summarize the theoretical basis of our finite size scaling 
analysis in section \ref{FSStheory}.
In section \ref{theAlgo} we discuss the Monte Carlo algorithm 
used in the simulations.
In section  \ref{Analysis}  we analyze the data and present 
the results for the fixed point values of the dimensionless ratios, inverse 
critical temperatures, the correction exponent $\omega$, and the 
critical exponents $\nu$ and $\eta$. Finally we  conclude and give an
outlook. 
In the appendix we discuss the dependence of the critical temperature 
and other non-universal quantities on $N$
and determine the RG-exponent $y_{N=6}$ related to a $Z_6$ invariant 
perturbation of the $O(2)$ invariant fixed point.

\section{The $(N+1)$-state clock model}
\label{theModel}
The model can be viewed as a generalization of the $N$-state clock model.  
The field $\vec{s}_x$ at the site $x=(x_0,x_1,x_2)$, where $x_i \in 0,1,2,...,L_i-1$,
 might assume one of the following values
\begin{equation}
\vec{s}_x \in
\left\{(0,0), \left(\cos(2 \pi m/N), \sin(2 \pi m/N)    \right)
\right\} \;,
\end{equation}
where $m \in \{1,...,N\}$. Compared with the $N$-state clock model, 
$(0,0)$ is added as possible value of the field variable. In our program, 
we store the field variables by using labels $m =0,1,2,...,N$. We assign
\begin{equation}
 \vec{s}(0) = (0,0)
\end{equation}
and for $m>0$
\begin{equation}
 \vec{s}(m) =  \left(\cos(2 \pi m/N), \sin(2 \pi m/N)   \right) \;\;.
\end{equation}
The reduced Hamiltonian is given by
\begin{equation}
\label{ddXY}
 {\cal H} = -  \beta \sum_{\left<xy\right>}  \vec{s}_x \cdot
     \vec{s}_y -D  \sum_x \vec{s}_x^{\,2} - \vec{H} \sum_x \vec{s}_x \;,
\end{equation}
where $\left<xy\right>$ denotes a pair of nearest neighbor sites on the simple cubic lattice.
We introduce the weight factor
\begin{equation}
\label{weight}
w(\vec{s}_x)  = \delta_{0,\vec{s}_x^{\,2}} +  \frac{1}{N} \delta_{1,\vec{s}_x^{\,2}}
              = \delta_{0,m_x} + \frac{1}{N} \sum_{n=1}^N \delta_{n,m_x}
\end{equation}
that gives equal weight to $(0,0)$ and the collection of all values $|\vec{s}_x| =1$.
Now the partition function can be written as 
\begin{equation}
 Z = \sum_{\{\vec{s}\} }  \prod_x w(\vec{s}_x) \; \exp(-{\cal H}) \;,
\end{equation}
where $\{\vec{s}\}$ denotes a configuration of the field.  
Note that in the limit $N \rightarrow \infty$, we recover the dynamically 
diluted
XY (ddXY) model studied in refs. \cite{XY1,XY2}. The reduced Hamiltonian 
of the ddXY model has the same form as eq.~(\ref{ddXY}):
\begin{equation}
\label{ddXY2}
{\cal H}_{ddXY} =  -  \beta \sum_{\left<xy\right>}  \vec{\phi}_x \cdot
\vec{\phi}_y - D  \sum_x \vec{\phi}_x^{\,2} - \vec{H} \sum_x \vec{\phi}_x \;,
\label{ddxy}
\end{equation}
where $\vec{\phi}_x$ is a vector with two real components.
The partition function is given by
\begin{equation}
 Z = \prod_x \left[\int d\mu(\phi_x) \right] \; \exp(- {\cal H}_{ddXY})  \;,
\end{equation}
with the local measure
\begin{equation}
\label{measure}
d\mu(\phi_x) =  d \phi_x^{(1)} \, d \phi_x^{(2)} \,
\left[
\delta(\phi_x^{(1)}) \, \delta(\phi_x^{(2)})
 + \frac{1}{2 \pi} \, \delta(1-|\vec{\phi}_x|)
\right] \; .
\label{lmeasure}
\end{equation}
Note that the dynamically diluted XY model is a special case ($K=0$) of the 
vectorialized Blume, Emery, and Griffiths (VBEG)  model studied in 
ref. \cite{MaKrDi04}.

\subsection{Phase diagram of the dynamically diluted XY model}
We expect that the phase diagram for  $N \ge 5$ is essentially the 
same as that of the ddXY model.
Therefore we briefly recall the results
obtained in refs. \cite{XY1,XY2}. In the limit $D \rightarrow \infty$ the 
XY model is recovered. There is a line of second order phase transitions
that ends at $D_{tri}$ in a tricritical point. 
Following ref. \cite{XY1}, based on mean-field calculations, $D_{tri}<0$.
Along the line of second order phase transitions, there is a $D^*$, 
where leading corrections to scaling vanish. We refer to the 
ddXY model at $D \approx D^*=1.06(2)$, ref. \cite{XY2},
as improved ddXY model.   In table \ref{lineddXY}
we summarize results obtained for the inverse critical temperature
$\beta_c$ at various values of $D$. 
\begin{table}
\caption{\sl \label{lineddXY}
Results for the inverse of the critical temperature $\beta_c$
for the dynamically diluted XY model. These results are taken 
from table II of ref. \cite{XY2}. 
}
\begin{center}
\begin{tabular}{ll}
\hline
\multicolumn{1}{c}{$D$}  & \multicolumn{1}{c}{$\beta_c$} \\
\hline
0.9  & 0.5764582(15)[9] \\
1.02 & 0.5637963(2)[2] \\
1.03 & 0.5627975(7)[7] \\
1.2 & 0.5470376(17)[6] \\
$\infty$ & 0.4541652(5)[6] \\
\hline
\end{tabular}
\end{center}
\end{table}
In the Appendix \ref{Ndependence} we shall study the $N$-dependence 
of $\beta_c$ in detail.

\subsection{Definitions of the measured quantities}
\label{def}
The quantities studied are essentially the same as in \cite{XY2}.
For completeness we list them below:
The energy density is defined as
\begin{equation}
\label{energy}
E=\frac{1}{V} \sum_{\left<xy\right>} \vec{s}_x \cdot \vec{s}_y\; .
\end{equation}
The magnetic susceptibility $\chi$ for a vanishing magnetization 
and the second moment correlation length $\xi_{2nd}$ are defined as
\begin{equation}
\label{chidef}
\chi  =  \frac{1}{V} \,
\biggl\langle \Big(\sum_x \vec{s}_x \Big)^2 \biggr\rangle
\end{equation}
and
\begin{equation}
\xi_{2nd}  =  \sqrt{\frac{\chi/F-1}{4 \sin^2 \pi/L}} \;,
\label{xidef}
\end{equation}
where
\begin{equation}
F  =  \frac{1}{V} \, \biggl\langle
\Big|\sum_x \exp\left(i \frac{2 \pi x_1}{L} \right)
         \vec{s}_x \Big|^2
\biggr\rangle
\end{equation}
is the Fourier transform of the correlation function at the lowest
non-zero momentum.
We consider several dimensionless quantities, which are also called
phenomenological couplings.
These quantities are, in the critical limit, invariant under RG
transformations. We
consider the Binder cumulant $U_4$ and its sixth-order generalization
$U_6$, defined as
\begin{equation}
U_{2j} = \frac{\langle(\vec{m}^2)^j\rangle}{\langle\vec{m}^2\rangle^j} \;,
\end{equation}
where $\vec{m} = \frac{1}{V} \, \sum_x \vec{s}_x$ is the magnetization of
the system.  We also consider the ratio $R_Z = Z_a/Z_p$ of the partition
function $Z_a$ of a system with anti-periodic boundary conditions in one of the
three directions and the partition function $Z_p$ of a system with periodic
boundary conditions in all directions.  Anti-periodic boundary conditions in
$0$-direction are obtained by changing the sign of the term
$\vec{s}_x \cdot \vec{s}_y$ of the Hamiltonian for links
$\left<xy\right>$ that connect the boundaries, i.e., for $x=(L,x_1,x_2)$ and
$y=(0,x_1,x_2)$. In order to avoid microscopic effects at the boundary, 
we require that $-\vec{s}_x$ is in the same set of values as $\vec{s}_x$. 
Therefore in the main part of the study $N$ is chosen to be even.
In the following we will refer to dimensionless ratios by $R$.
Derivatives of dimensionless ratios with respect to the 
inverse temperature
\begin{equation}
S_R = \frac{\partial R}{\partial \beta}
\label{sdef}
\end{equation}
are used to determine the critical exponent $\nu$. In the following these
quantities are also denoted by slope of $R$.

For most of our analysis we need the observables as a function of 
$\beta$ in a certain neighborhood of the critical point.  
To this end, we simulate at $\beta_s$,
which is a good approximation of $\beta_c$. In order to extrapolate
in $\beta$ we compute the coefficients of the Taylor series in $\beta-\beta_s$
for all quantities listed above up to the third order. Note that a
reweighting analysis is not possible, since, due to the large statistics, 
we performed a binning of the data already during the simulation.

\section{Finite size scaling:  theoretical background}
\label{FSStheory}
The account given below is similar to section II B of ref. \cite{XY1}. 
The main purpose is to make the present paper self contained. Our assumptions
concerning subleading corrections differ from  ref. \cite{XY1}. See section
\ref{secirr} below.
Our starting point is the finite size scaling behavior of the reduced free
energy density, which is defined by
\begin{equation}
f(\beta,h,D,L) = - \frac{1}{V}  \ln Z(\beta,h,D,L)  \;,
\end{equation}
where $Z$ is the partition function and $V=L^3$ is the number of lattice sites.
Note that there is also a dependence on $N$ that we suppress in the following
to keep the notation tractable.

The reduced free energy density can be written in terms of the analytic
functions ${\cal F}_{sing}$ and $g$, see for example eq.~(2.14) of ref. 
\cite{PeVi},
\begin{equation}
\label{freeenergy}
 f(\beta,h,D,L) = L^{-d} {\cal F }_{sing}(L^{y_t} u_t, L^{y_h} u_h, \{u_i L^{y_i}\})
 + g(\beta,h,D) \;,
\end{equation} 
where $d$ is the dimension of the system.
Note that ${\cal F}_{sing}$ is a universal function, which
however depends on the global geometry of the system, for example on 
aspect ratios $L_i/L_j$, where $i \ne j$ are the directions on the lattice or
on the type of boundary conditions. Here we consider periodic and anti-periodic
boundary conditions that do not generate boundary contributions
like Dirichlet boundary conditions for example. The analytic background
$g(\beta,h,D)$ does not depend on these global properties. $u_t$ and $u_h$ are the 
temperature like and external field like scaling fields  with the 
RG-exponents $y_t$ and $y_h$, respectively. These are the only
relevant RG-exponents: $y_t>0$ and $y_h>0$.  In addition  there are
irrelevant RG-exponents $y_i < 0$.  Below we summarize results on irrelevant 
RG-exponents given in the literature. Following for example 
ref. \cite{PeVi}, section 1.5.7, the non-linear scaling fields can be
written as  
\begin{eqnarray}
u_t &=& g_{01}(D) \; t +  g_{11}(D) \; t^2 + g_{12}(D) \; h^2 +O(t^3, t h^2,h^4) \; ,\\
u_h &=& g_{02}(D) \; h \left[ 1 + g_{12}(D) \; t + g_{22}(D) \; h^2   +O(t^2, t h^2,h^4)
    \right]  \; ,
\end{eqnarray}
where we define the reduced temperature as $t= \beta_c(D) - \beta$.
Note that $\beta_c(D)$ and the coefficients $g_{ij}(D)$ depend on $N$. 
In appendix \ref{Ndependence} we show however that there is a fast
convergence as $N \rightarrow \infty$.
The external field is written as $\vec{H} =  h \vec{H}_0$, where $\vec{H}_0$
is a two-component unit vector. 
We have introduced $g_{01}(D)$ and $g_{02}(D)$
to get the same function ${\cal F}_{sing}$ for all values of $D$ 
on the critical line.
The scaling field of the leading correction is
\begin{equation}
\label{udrei}
u_3 = g_{13}(D) + g_{23}(D) \; t + g_{33}(D) \; h^2 + O(t^2, t h^2,h^4) \;.
\end{equation}
The improved model is characterized by $g_{13}(D^*) = 0$.  
Note that in general $g_{23}(D^*) \ne 0$ and $g_{33}(D^*) \ne 0$. 
Also note that $D^*$ depends on $N$, since $g_{13}(D)$ depends on $N$.
For numerical results see appendix \ref{DsonN}.

\subsection{Irrelevant RG-exponents}
\label{secirr}
Let us briefly summarize results on RG-exponents for the three-dimensional
XY-universality class given in the literature.
Various methods give a, at least qualitatively, consistent picture for the 
relevant RG-eigenvalues $y_t$ and $y_h$ and the leading irrelevant RG-eigenvalue 
$y_3$. Using scaling relations, see for example ref. \cite{PeVi}, sects. 1.3 and 1.5.1, 
these are related with the critical exponents given in table \ref{methods} as
\begin{equation}
 y_t  =  1/\nu \;, \; \;\;
 y_h  =  \frac{d + 2 -\eta}{2}  \;, \; \;\;
 y_3  =  -\omega  \;.
\end{equation}
Scaling fields can be classified according to the symmetry properties 
of the operators associated to them. The simple cubic lattice breaks the 
Galilean symmetries of continuous space. The leading correction associated
has the RG-exponent $y_{NR} =-2.02(1)$ \cite{ROT98,XY1,XY2}. Note that in the
case of the three-dimensional Ising universality class,  $y_{NR}=-2.0208(12)$
given in table I of ref. \cite{Campostrini:2002cf}  
is in reasonable agreement with $y_{NR}=-2.022665(28)$ that
follows from $\Delta=5.022665(28)$ for angular momentum $l=4$ given in
table 2 of ref. \cite{Simmons-Duffin:2016wlq}.

Results for subleading corrections are provided by different incarnations
of the renormalization group. 
Newman and Riedel \cite{NewmanRiedel} studied the fixed point of the
 $O(N)$ invariant 
$\phi^4$ theory in three dimensions using the scaling field method.
They predict by using the scaling field method subleading corrections
with $y_{421} = -1.77(7)$  and $y_{422} = -1.79(7)$,
which are nearly degenerate. For the meaning of the indices see 
ref. \cite{NewmanRiedel}. In refs. \cite{XY1,XY2} the analysis of the data
is based on this result.  Note that Newman and Riedel find $y_{422} = -1.67(11)$
in the case of the Ising universality class, which is not confirmed by 
the conformal bootstrap method. Instead, $y''=-3.8956(43)$ is found, see 
the estimate related to the operator $\epsilon''$ given in table 2 of 
ref. \cite{Simmons-Duffin:2016wlq}.
In fact, the estimates for subleading correction exponents obtained by the 
functional renormalization group (FRG), see for example ref. \cite{Litim04},
are in better agreement with those of the conformal bootstrap method.
In table 3 of ref. \cite{Litim17} results for correction exponents for a 
large range of $N$, where $N$ refers to the $O(N)$ symmetry of the theory,
are given. The qualitative picture is the same for all $N$ and
the numerical values change slowly with varying $N$. Therefore we regard
it as plausible that $-3.5 \gtrapprox  y'' \gtrapprox  -4$ for the 
three-dimensional XY-universality class.  Note that skipping corrections
$\propto L^{-1.77}$ in the analysis of our data virtually does not 
change the central values of the final results. Estimates of the error
are reduced by a factor of $\approx 2/3$. 

Finally let us recall the results for the RG-exponent  associated with
a $Z_N$ invariant perturbation. The authors of ref. \cite{Debasish} 
obtain $-y_N= 0.128(6)$, $1.265(6)$, $2.509(7)$, $3.841(8)$, $5.278(9)$, 
$6.796(9)$, $8.399(10)$, $10.077(11)$, and  $11.825(12)$ for $N=4$, $5$,
$6$, ..., $12$, respectively. In the main part of our study we have
simulated the $(N+1)$-state clock model for $N=8$. For this value of $N$, 
we can ignore deviations from $O(2)$-invariance in the finite size scaling 
analysis of our data as can be clearly seen from the analysis presented in
appendix \ref{allonN}.

\subsection{The magnetic susceptibility and the energy density}
The magnetic susceptibility at $h=0$ for vanishing magnetization is
\begin{equation}
\chi= - \frac{2}{V} \left . \frac{\partial^2 f}{\partial h^2} \right |_{h=0} = 
\frac{1}{V} \left \langle \left( \sum_x \vec{s}_x \right)^2 \right\rangle \;\;.
\end{equation}
Note that we have introduced a factor of two here, to stay consistent with the
definition~(\ref{chidef}) above.

Let us define $\tilde u_t =  u_t L^{y_t}$, $\tilde u_h =  u_h L^{y_h}$, and
$\tilde u_i =  u_i L^{y_i}$.
Now let us compute the second partial derivative of $f$ with respect to $h$ at
$h=0$:
\begin{equation}
\left . \frac{\partial^2 f}{\partial h^2} \right |_{h=0} =
L^{-d} \left . \frac{\partial^2 {\cal F}_{sing}}{\partial h^2} \right|_{h=0}
+ \left . \frac{\partial^2 g}{\partial h^2} \right|_{h=0}  \;,
\end{equation}
where
\begin{eqnarray}
L^{-d} \left . \frac{\partial^2 {\cal F}_{sing}}{\partial h^2} \right|_{h=0}
& = & \left . \frac{\partial {\cal F}_{sing}}{\partial \tilde u_t} \right|_{h=0}
    2 \; (g_{12}(D) + ...) \; L^{y_t-d}   \nonumber \\
& + &
\left . \frac{\partial^2 {\cal F}_{sing}}{\partial \tilde u_h^2} \right|_{h=0}
 \left (g_{02}(D) \;  \left[ 1 + g_{12}(D) \; t + ...\right] \right)^2 
   L^{2 y_h-d} + ... \;.
\label{chiF}
\end{eqnarray}
There are also contributions stemming from partial derivatives with respect 
to $\tilde u_i$. However these are related with correction exponents 
$\epsilon > 4$ and therefore play little role in the analysis of the data. 

It remains to Taylor expand 
$\left . \frac{\partial^2 {\cal F}_{sing}}{\partial \tilde u_h^2} \right|_{h=0}$ 
and $\left . \frac{\partial {\cal F}_{sing}}{\partial \tilde u_t} \right|_{h=0}$
in $\tilde u_i$. We arrive at corrections that are proportional to $L^{y_3}$,
$L^{2 y_3}$, $L^{3 y_3}$, ..., $L^{y_{NR}}$, $L^{y_{NR}+y_3}$, ..., $L^{y''}, ...$.
Note that for an improved model, all terms with $y_3$ in the exponent have a 
vanishing amplitude, since $u_3=0$.
For an improved model, at the critical point we get
\begin{equation}
\label{chic}
 \chi_{h=0,t=0, D=D^*} = 
a L^{2 y_h-d} \; \left[1+c_{NR}  L^{y_{NR}} + c'' L^{y''} + c_t L^{y_t-2 y_h} + ... \right ]  + b \;.
\end{equation}
Note that $2 y_h-d = 2-\eta$.  The analytic background $b$ can be viewed as a correction with the
RG-exponent $y_b=\eta-2 \approx -1.962$, which is close to $y_{NR}=-2.02(1)$.  
Also the value of $y_t-2 y_h \approx -3.473$ is close to that of $y''$. 

The energy density, eq.~(\ref{energy}), is given by the first derivative of the free energy 
with respect to $t$. At the critical point we get
\begin{equation}
\left . \frac{\partial f}{\partial t} \right|_{t=0,h=0} =  
\left . \frac{\partial {\cal F}_{sing}}{\partial \tilde u_t} \right|_{t=0,h=0}
   g_{01}(D) L^{y_t-d} 
 + \left . \frac{\partial {\cal F}_{sing}}{\partial \tilde u_3} \right|_{t=0,h=0} g_{23}(D) L^{y_i-d}  + 
\left . \frac{\partial g}{\partial t} \right|_{t=0,h=0}  \;.
\end{equation}
It remains to Taylor expand $\left . \frac{\partial {\cal F}_{sing}}{\partial \tilde u_t} \right|_{t=0,h=0}$
in $\tilde u_i$.  We arrive at
\begin{equation}
\label{enescaling}
 E = E_0 + a L^{y_t-d} \; \left(1 + c_{NR}  L^{y_{NR}} + c_3 L^{y_3- y_t} +  c'' L^{y''} + ...   \right)
\end{equation}
for an improved model at the critical point. Note that $y_3- y_t \approx -2.278$ is only slightly 
smaller than $y_{NR}$.

\subsection{Phenomenological Couplings}
A cornerstone of our analysis are dimensionless quantities which are also 
called phenomenological couplings. In the following we shall denote them by $R$,
since in our case they are ratios. The first quantity that we consider is 
the ratio of partition functions. We get
\begin{equation}
\ln \frac{Z_a}{Z_p} = V (f_p - f_a) = {\cal F}_{p,sing}  - {\cal F}_{a,sing} \;,
\end{equation}
since the analytic background exactly cancels. Hence
\begin{equation}
\label{scalingR}
 \frac{Z_a}{Z_p} = R_Z(L^{y_t} u_t,L^{y_h} u_h,\left\{L^{y_j} u_j \right\}) \;\;.
\end{equation}
In addition we study the cumulants
\begin{equation}
 U_{2j} = \frac{\langle m^{2 j} \rangle}{\langle m^{2} \rangle^j}
\end{equation}
for $j=2$ and $3$.  Here we can build on the result obtained above for the 
magnetic susceptibility. Also $\langle m^{2 j} \rangle$ can be computed
from partial derivatives of the free energy density with respect to 
the external field $h$. The dominant contributions stem from the derivatives
of the singular part of the free energy with respect to $\tilde u_h$
and even derivatives of 
the singular part of the free energy with respect to $\tilde u_t$. 
Hence 
\begin{equation}
\label{Uexpansion}
 U_{2j} =  R_U(L^{y_t} u_t,L^{y_h} u_h,\left\{L^{y_i} u_i \right\}) + a L^{-2 y_h +d}
+ b L^{-2 y_h +y_t} + ... \; .
\end{equation}
In the case of the second moment correlation length $\xi_{2nd}$ divided by the 
linear lattice size 
$L$ we also expect corrections that go back to the magnetic susceptibility. 
In addition there 
is a correction $\propto L^{-2}$ due to the construction of $\xi_{2nd}$.

Taking the derivative of a phenomenological coupling with respect
to the reduced temperature $t$ we get
\begin{equation}
\left . \frac{\partial R}{\partial t} \right |_{h=0} = \left . \frac{\partial R}{\partial \tilde u_t} \right |_{h=0}
\; (g_{01}(D) + g_{11}(D) t + ... )  L^{y_t}  
+ \left . \frac{\partial R}{\partial \tilde u_3} \right |_{h=0} 
 g_{23}(D) L^{y_3} + ... \;.
\label{slopeF}
\end{equation}

At the critical point of an improved model
\begin{equation}
\left . \frac{\partial R}{\partial t} \right |_{t=0, h=0, D=D^*}
 =  a L^{y_t} \; 
\left(1 + c L^{y_{NR}} + ... + d \; g_{23} L^{-y_t+y_3} + ... \right) \; ,
\label{slopeFC}
\end{equation}
where we performed a Taylor expansion of 
$\frac{\partial R}{\partial \tilde u_t}$  and
$\frac{\partial R}{\partial \tilde u_3}$ 
 with respect to $\{\tilde u_i \}$.

\subsection{Fixing the value of $R$}
In the analysis of our data, we consider certain quantities at a fixed 
value $R_{f}$ of a dimensionless quantity. This means that for each 
lattice size $L$, we compute $\beta_{f}(D,L)$ defined by
\begin{equation}
R(\beta_{f}(D,L),D,L) = R_{f} \; .
\end{equation}
Note that we have skipped the argument $h$, since $h=0$ throughout. Making use of eq.~(\ref{scalingR}) we
get
\begin{equation}
R(\beta_{f},D,L) = R^* + a(D) (\beta_c(D)-\beta) L^{y_t} + ...  + c(D) L^{y_3} + ...  \;\;.
\end{equation}
for $R_f \approx R^*$, where $R^*$ is the fixed point value of $R$. 
Hence 
\begin{equation}
\label{betaf}
 \beta_{f}(D,L) =  \beta_c(D) - a(D)^{-1} (R^*- R_{f}) L^{-y_t} + ... + 
                  a(D)^{-1} c(D) L^{y_3-y_t} +  a(D)^{-1} d(D) L^{y_{NR}-y_t} + ...  \;.
\end{equation}
Note that $c(D^*)=0$. 
First we consider a phenomenological coupling $R_2$ at a fixed value $R_{1,f}$ of an
other phenomenological coupling $R_1$. One gets 
\begin{equation}
\label{RRfix}
 R_2(R_{1,f},D,L)  = r_2(R_{1,f},\{\tilde u_i\}) + c(R_{1,f},D) L^{y_3-y_t} + ...+ d(D) L^{2 y_3-y_t} + ...\; ,
\end{equation}
where $ c(R_{1}^*,D) =0$. Note that the corrections are due to the fact that the $u_i$  depend 
on $t$, see eq.~(\ref{udrei}).

We also compute the magnetic susceptibility and the slope of 
phenomenological couplings at $R_f$. 
Plugging in eq.~(\ref{betaf}) into eqs.~(\ref{chiF}, \ref{slopeF}) we see 
that compared with eqs.~(\ref{chic},\ref{slopeFC}) additional correction
terms proportional to $(R_f-R^*) L^{-y_t}$, $(D-D^*) L^{-y_t+y_3}$ and
$L^{-y_t+y_{NR}}$  appear. Therefore it is favorable to take 
$R_f \approx R^*$. In the numerical analysis, one should vary $R_f$ 
to check the effect of a possible deviation from $R^*$. 

\section{The algorithm}
\label{theAlgo}
As in previous studies, for example  refs. \cite{XY1,XY2}, we have implemented 
a hybrid of local Metropolis updates, the single cluster update \cite{Wolff}, 
and the wall cluster update \cite{KlausStefano}.  Now let us discuss in detail these
components of the algorithm and their implementation.

\subsection{Local Metropolis algorithm}
\label{localAlg}
As usual, in the elementary step of the local update, the variable at a single 
site is changed, while all other variables are kept fixed. Using these 
elementary updates, we go through the lattice in typewriter fashion.  Going
through the lattice once is called a sweep. 
We use two different ways to generate the proposal for the local Metropolis
update. In both cases, the proposal $\{\vec{s}\}'$ is accepted with the 
probability 
\begin{equation}
	P_{acc} = \mbox{min}[1,\exp(-\Delta H)] \;,
\end{equation}
where 
\begin{equation}
\Delta H =  H(\{\vec{s}\}') - H(\{\vec{s}\}) \;.
\end{equation}
The weight, eq.~(\ref{weight}), is taken into account by the probabilities 
used to generate the proposal.
The first choice is given by the following probabilities:
If $\vec{s}_x = (0,0)$ we take with equal probability one of the 
$N$ values with  $|\pvec{s}_x'| = 1$ as proposal.  Else, for
$|\vec{s}_x| = 1$,  we always take $\pvec{s}_x' = (0,0)$ as proposal.  

For an efficient implementation, one should avoid to compute $\exp(.)$ 
for each update step. Instead we should store possible results in a table 
before the actual simulation is started.

The sum of all nearest neighbor spins can take a too large number of 
possible values to store $\exp(-\Delta H)$ efficiently. 
Therefore we tabulate instead the contribution to the Boltzmann factor 
by pairs
\begin{equation}
 B(m,n) = \exp(\beta \; \vec{s}(m) \cdot \vec{s}(n) )
\end{equation}
and its inverse $B^{-1}(m,n)$. Furthermore $\exp(-D)$ and $\exp(D)$ are
computed once and are then stored.
Then, for $m_x=0$, where $x$ is the site to be updated, we get 
\begin{equation}
\exp(-\Delta H) =\exp(D) \; \prod_{y.nn.x} B(m_x',m_y) \;, 
\end{equation}
where the product runs over all nearest neighbors (nn) of $x$. 
Note that $B(0,n)=1$ for all values of $n$. For $m_x>0$ we get
\begin{equation}
\exp(-\Delta H) =\exp(-D) \; \prod_{y.nn.x} B^{-1}(m_x,m_y) \; . 
\end{equation}

Since we were not able to prove the ergodicity of this algorithm, 
we used in addition a second choice of the proposal.
It is generated independently of the old value of the variable.
With probability $1/2$ we take $\pvec{s}_x' = (0,0)$ and  with 
equal probabilities $1/(2 N)$ one of the remaining values is chosen.
Here
\begin{equation}
\exp(-\Delta H) =\exp(-D [\vec{s}_x^{\,2}-\pvec{s}_x'^{\,2}]) 
\;\prod_{y.nn.x} [B^{-1}(m_x,m_y) B(m_x',m_y)  ]\; .
\end{equation}
This update costs more CPU time than the first. However ergodicity
is obvious.

\subsection{Cluster algorithms}
Cluster algorithms can be applied without major modifications compared with 
the ddXY model. 
We just have to note that the reflection has to be chosen such that the 
field variables remain in the allowed set of values. A reflection is given by
\begin{equation}
 \vec{s}\,' = \vec{s} - 2 ( \vec{r} \cdot \vec{s} \, ) \vec{r} \;,
\end{equation}
where
\begin{equation}
\vec{r}  =  \left(\cos(\pi m/N), \sin(\pi m/N) \right)
\end{equation}
with $m=0, 1, 2, ...,N-1$. The cluster update is characterized by the delete probability
\begin{equation}
p_d(\vec{s}_x, \vec{s}_y) = \mbox{min}\left[1, \exp\left(-2 \beta [\vec{r} \cdot \vec{s}_x][\vec{r} \cdot \vec{s}_y]\right) \right]
\; .
\end{equation}
The values of $p_d$ are tabulated before the actual simulation is started. 
For a discussion of the 
single cluster \cite{Wolff} and the wall cluster update \cite{KlausStefano} 
used for the simulation of the ddXY model see \cite{XY1,XY2}.

\subsection{The implementation}
Our simulations are organized in a similar fashion as in \cite{XY1,XY2}.
Since we could not store the results of all measurements on hard disc, we performed
a binning of the data during the simulation. 

During the study we varied the precise composition of the update cycle.
In most of the simulations the following cycle, given by a C-code, is used:
\begin{verbatim}
for(i=0;i<N_bin;i++)
  {
  Metropolis_2(); 
  for(k=0;k<6;k++)
    {
    Metropolis_1();
    for(j=0;j<L;j++) single_cluster();
    Metropolis_1();
    wall_cluster(direction=k%3);
    measurements();
    }
  }
\end{verbatim}
Here \verb+Metropolis_1()+ and \verb+Metropolis_2()+ are sweeps, using 
the first and second type of the Metropolis update discussed
in section \ref{localAlg}.
The single cluster update is given by  \verb+single_cluster()+ and
\verb+wall_cluster(direction=k%3)+ is a wall cluster update for one 
of the three spacial  directions.  The plane is perpendicular to the 
$k$-axis.  The position of the plane is randomly chosen in 
$\{0, 1, 2, ...,L-1\}$. In order to compute $Z_a/Z_p$ we need two 
subsequent wall cluster updates, where the two reflection axes are
perpendicular.
The first axis is chosen randomly among the $N$ possible directions.

We did run our program on standard x86 CPUs.  For lack of human time,
we made no attempt to implement our program on a graphics processing unit (GPU). 
For cluster algorithms on GPUs see for example refs. \cite{Weigel,Komura}.

Let us briefly comment on the CPU time required by the different components
of the update cycle. We performed the simulations on various PCs and servers
at the institute of theoretical physics. Here we quote numbers for a single
core of an Intel(R) Xeon(R) CPU E3-1225 v3 running at 3.20 GHz.
We implemented the code in standard C and used the SIMD-oriented Fast Mersenne
Twister algorithm \cite{twister} as random number generator. 

Our Metropolis update type one requires $1.2 \times 10^{-8}$ s per site.  In the case 
of the single cluster update about  $3.8 \times 10^{-8}$ s per site are needed.
Note that the random number generator requires for one sequential access about
$3  \times 10^{-9}$ s.  Compared with our program for the ddXY model,
these updates are faster by roughly a factor of three.

Plots were generated by using the Matplotlib library \cite{plotting}. The least 
square fits were performed by using the function {\sl curve\_fit()} contained in the 
SciPy  library \cite{pythonSciPy} with the default Levenberg-Marquardt algorithm 
\cite{LM1,LM2,LM3}. The  function {\sl curve\_fit()} acts as a wrapper to functions
contained in the MINPACK library \cite{MINPACK}.

\section{The simulations and the analysis of the data}
\label{Analysis}
\begin{figure}
\begin{center}
\includegraphics[width=14.5cm]{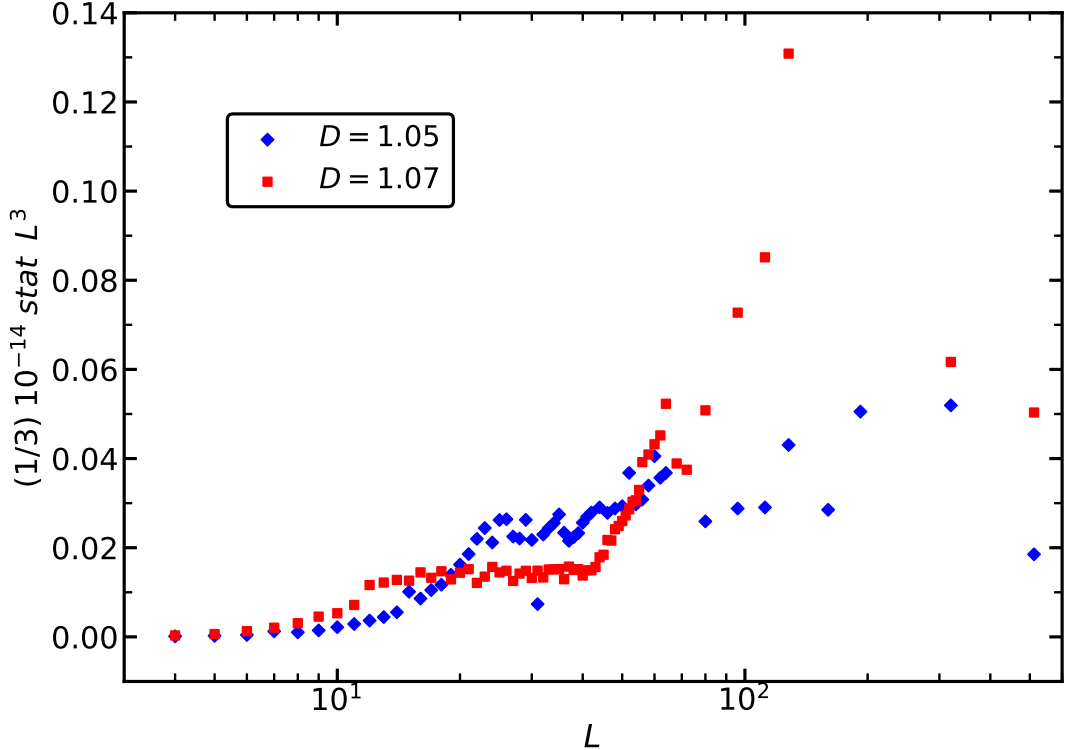}
\caption{\label{stat}
We plot the number of measurements times the volume $L^3$ as a function 
of the linear lattice size $L$ for $N=8$ at $D=1.05$ and $D=1.07$.
}
\end{center}
\end{figure}

We simulated the model for $N=8$ at various values of $D$, close to the 
inverse critical temperature $\beta_c(D)$.  Most CPU time is spend on 
simulations for  $D=1.02$, $1.05$, and $1.07$ which are close to $D^*$.
We simulated linear lattice sizes up to $L=512$, where the 
statistics is decreasing with increasing $L$. 
In figure \ref{stat} we plot the number of measurements times the volume $L^3$ 
as a function of the linear lattice size $L$ for $D=1.05$ and $1.07$.
In the case of $D=1.02$ the statistics is similar but we have simulated at 
fewer lattice sizes in the range $L=20$ up to $80$.
In addition we simulated at $D=-0.7$, $-0.5$, $0$, $0.45$, $0.9$, $1.24$, 
and $\infty$.  In these cases we considered linear lattice sizes up to
$L=64$, $64$, $72$, $72$, $48$, $48$, and $72$, respectively.
The main purpose of these simulations is to determine the correction 
exponent $\omega$.
A few simulations at $D=-0.85$, $-0.86$, and $-0.87$ are performed to obtain a rough 
estimate of the tricritical point. 

Our simulations were performed on various PCs and servers.
 In total these simulations took the equivalent 
of about 50 years of CPU time  on a single core of a
Intel(R) Xeon(R) CPU E3-1225 v3 running at 3.20 GHz.
Note that the study was not systematically
designed at the start but grew with time, also depending on the availability
of CPU time.

Let us briefly comment on the assessment of the error of the final estimates 
for critical exponents and other quantities of interest.
In the ans\"atze we can take into account only a small number of correction 
terms. This inevitably leads to systematic errors caused by corrections
to scaling that are not explicitly taken into account.
A large $\chi^2/$d.o.f. indicates that the ansatz is not adequate to
represent the data. However, dealing with an ansatz that not fully
represents the underlying function, a small $\chi^2/$d.o.f. and a 
corresponding 
acceptable goodness of the fit says very little on the deviation of 
the fit parameters from their true values. 
In order to get some handle on systematic errors caused by corrections to
scaling
that are not taken into account in the ansatz, we either consider a number
of different quantities or ans\"atze with a different number of correction 
terms. 
The final estimate and its error bar are chosen such that these
 different estimates are covered. The actual choice, which
fits and minimal lattice sizes are taken into account, is at least
partially an ad hoc decision. To allow the reader an own assessment, the 
direct outcome of fits is given in figures. We made no effort to give
a separate estimate of the statistical and systematical error, since they
are interwoven in our assessment.

The analysis of the data is organized in the following way:
First we perform joint fits of our data for the dimensionless quantities
$R$ for $D=1.02$, $1.05$ and $1.07$.  The results are the fixed 
point values $R^*$ and estimates of the inverse critical temperatures.
Next we include values of $D$ with a larger amplitude of the leading 
correction to determine the exponent $\omega$. To this end we analyze
the cumulants $U_4$ and $U_6$ at a fixed values of either $Z_a/Z_p$ or
$\xi_{2nd}/L$. Then we determine $D^*$ focusing again on 
$D=1.02$, $1.05$ and $1.07$. It follows a rough localization of the 
tricritical point $D^*$. 
In the final step of the analysis, we determine the critical exponents
$\nu$ and $\eta$. To this end we analyze the finite size scaling behavior  
the slopes of dimensionless quantities $R$, the energy density and the 
magnetic susceptibility. 

\subsection{The critical coupling $\beta_c$ and the fixed point values of 
dimensionless ratios $R^*$}
\label{betac0}
First we determined the critical coupling $\beta_c(D)$ and the fixed point
values $R^*$ of the dimensionless quantities that we have computed.
To this end we analyzed our data at $D=1.02$, $1.05$, and  $1.07$, which are close to $D^*$. 

Motivated by eqs.~(\ref{scalingR},\ref{Uexpansion}), we have fitted our 
data with  four different ans\"atze
\begin{eqnarray}
  \label{xxxx0}
 R(L,D,\beta_c(D)) &=& R^* \;\;, \\
  \label{xxxx1}
 R(L,D,\beta_c(D)) &=& R^* + b(D) L^{-\epsilon_1} \;\;, \\
  \label{xxxx2}
 R(L,D,\beta_c(D)) &=& R^* + b(D) L^{-\epsilon_1} + c(D) L^{-\epsilon_2}  \;\;, \\
  \label{xxxx3}
 R(L,D,\beta_c(D)) &=& R^* + b(D) L^{-\epsilon_1} + c(D) L^{-\epsilon_2} + d(D) L^{-\epsilon_3} \;\;.
\end{eqnarray}
We need the phenomenological couplings $R$ as a function of 
the inverse
temperature. To this end we have used the Taylor series around
the value $\beta_s$ of the inverse temperature used in the
simulation. We have checked that $\beta_c$ and $\beta_s$
are sufficiently close to avoid significant truncation effects.
This way, for example eq.~(\ref{xxxx1}) becomes
\begin{equation}
 R(L,\beta_s) = R^* - c_1(L,\beta_s) (\beta_c-\beta_s)
                    - \frac{c_2(L,\beta_s)}{2!} (\beta_c-\beta_s)^2
                    - \frac{c_3(L,\beta_s)}{3!} (\beta_c-\beta_s)^3 \;\;,
\nonumber
\end{equation}
where $R^*$ and $\beta_c$ are the two parameters of the fit.

It turned out that fits with the ansatz~(\ref{xxxx1})
are not very useful, since the amplitude of leading corrections is small
for the values of $D$ considered here. Therefore we shall not discuss 
the results of the these fits in the following. Furthermore we did not
consider ans\"atze with $\epsilon_2 = 2 \omega$ here, since the amplitude 
of such corrections should be very small. This will be verified below 
in section \ref{corrections}.
In the case of $Z_a/Z_p$ we have used in eq.~(\ref{xxxx2}) the choices 
$\epsilon_1=0.79$ and $\epsilon_2=2.02$. In eq.~(\ref{xxxx3}) we used
in addition either $\epsilon_3=3.5$ or $\epsilon_3=4$.
Note that below, in section \ref{corrections}, 
we shall find $\omega=0.789(4)$, eq.~(\ref{omegares}). 

We performed a preliminary analysis using different parameterizations and
choices of data sets. Based on this analysis we decided to extract 
the final results in the following way:
We performed joint fits for the three values $D=1.02$, $1.05$, and $1.07$, 
where we parameterize the amplitude of the leading correction as 
\begin{equation}
 b(D) = b_s (D - D^*)
\end{equation}
and the amplitudes of higher corrections, $c(D)$ and $d(D)$ 
are assumed to be the same for all three values of $D$.

\begin{figure}
\begin{center}
\includegraphics[width=14.5cm]{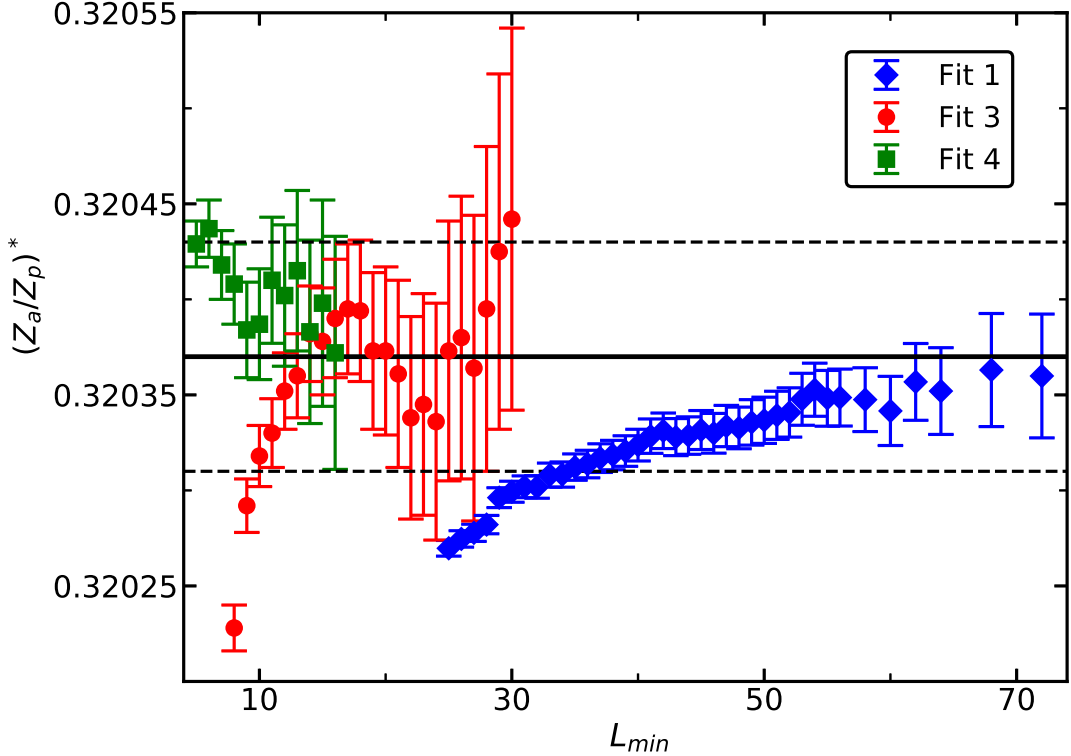}
\caption{\label{ZAZPstar}
We give the results for $(Z_a/Z_p)^*$ fitting with the ans\"atze 
(\ref{xxxx0},\ref{xxxx2}) and (\ref{xxxx3}) with $\epsilon_3=4$, 
corresponding to fit 1, 3, and 4 in the legend of the
figure, as a function of the minimal lattice
size $L_{min}$ that is included in the fit. Data for $D=1.02$, $1.05$
and $1.07$ are jointly fitted. The solid line gives our final estimate and the 
dashed ones the corresponding error.
}
\end{center}
\end{figure}

First we analyzed the data for the ratio of partition functions $Z_a/Z_p$.
In figure \ref{ZAZPstar} we plot results for $(Z_a/Z_p)^*$ of fits
using the ans\"atze~(\ref{xxxx0},\ref{xxxx2},\ref{xxxx3}).  
We give only data points with $\chi^2/$d.o.f.$ < 4$.  In the case
of ansatz~(\ref{xxxx0}) we see that $\chi^2/$d.o.f. 
decreases rapidly with increasing
$L_{min}$, where $L_{min}$ is the minimal lattice size that is included into 
the fit. For $L_{min}=33$, $\chi^2/$d.o.f. $= 1.012$ is reached. 
For ansatz~(\ref{xxxx2}) we find $\chi^2/$d.o.f. $=0.986$ already 
for $L_{min}=9$. As amplitude of the correction $\propto L^{-2.02}$ 
we find $c \approx -0.07$. For ansatz~(\ref{xxxx3}) with $\epsilon_3=4$ 
we find $\chi^2/$d.o.f. 
$=0.972$  for $L_{min}=5$. The amplitude of the correction $\propto L^{-4}$ is 
$d \approx -1.6$.  Consistently with ansatz~(\ref{xxxx2}) find 
$c \approx -0.06$.  Using $\epsilon_3=3.5$ instead, we get $\chi^2/$d.o.f.
$=1.135$  for $L_{min}=5$ and $\chi^2/$d.o.f. $=0.889$ for $L_{min}=7$.
For $L_{min}=7$ we get $d = -0.85(4)$ and $c = -0.028(5)$.
The fact that the amplitude of the 
correction $\propto L^{-\epsilon_3}$ is much larger than that of 
$\propto L^{-2.02}$ is surprising.

Our final estimate
\begin{equation}
 (Z_a/Z_p)^* = 0.32037(6)
\end{equation}
is taken such that it is consistent with the results of the three different
ans\"atze.
Note that we also varied the values of $\epsilon_1$ and $\epsilon_2$ within
the range of the expected error bars. The results of the fits change little.
In a similar way we arrive at the estimates for $D^*$ and $\beta_c$ at
$D=1.02$, $1.05$ and $1.07$.  These estimates are given in table
\ref{betac1}.

Next we analyzed the data for the ratio $\xi_{2nd}/L$ and the cumulants 
$U_4$ and $U_6$ in a similar way, taking into account that 
also corrections $\propto L^{\eta-2}$ might be present. 
The final results are summarized in table \ref{betac1}.

\begin{table}
\caption{\sl \label{betac1}
In the first column the phenomenological coupling is specified.
In the second column we give the corresponding 
estimates of the fixed point values $R^*$.
In the third column we give the estimates of $D^*$, where leading 
corrections to scaling vanish.  In the following columns, the estimates
of the inverse critical temperature $\beta_c$ for $D=1.02$, $1.05$, and 
$1.07$ are given. These estimates are based on joint fits of our data
for $D=1.02$, $1.05$, and $1.07$, as discussed in the text.
In the last row we give our final estimates of $\beta_c$. 
}
\begin{center}
\begin{tabular}{cllllll}
\hline
\multicolumn{1}{c}{$R$}&\multicolumn{1}{c}{$R^*$} & \multicolumn{1}{c}{$D^*$} & 
\multicolumn{1}{c}{$\beta_c(1.02)$} & \multicolumn{1}{c}{$\beta_c(1.05)$} & 
\multicolumn{1}{c}{$\beta_c(1.07)$} \\
\hline
$Z_a/Z_p$ & 0.32037(6)& 1.065(35)& 0.56379620(8)&0.56082390(7)&0.55888342(7)\\  
$\xi_{2nd}/L$&0.59238(7)&1.075(25)&0.56379622(9)&0.56082391(8)&0.55888342(8)\\
$U_4$       &1.24296(8) &1.054(10)&0.56379626(8)&0.56082386(8)&0.55888335(10)\\
$U_6$       &1.75040(25)&1.054(10)&0.56379626(8)&0.56082386(8)&0.55888335(10)\\
\hline
            &           &         &0.56379622(10)&0.56082390(10)&0.55888340(10)\\
\hline
\end{tabular}
\end{center}
\end{table}
The estimates for $R^*$ can be compared with $(Z_a/Z_p)^*=0.3203(1)[3]$, 
$(\xi_{2nd}/L)^*=0.5924(1)[3]$, $U_4^*=1.2431(1)[1]$, 
and $U_6^*=1.7509(2)[7]$ given in table I of \cite{XY2}. These results
were obtained by analyzing data obtained for the 2-component $\phi^4$ 
and the ddXY model on the simple cubic lattice. In ref. \cite{XY2}
the authors tried to distinguish between statistical $()$ and systematical $[]$
error. We find a nice agreement of the estimates, giving support to the 
hypothesis that the improved (8+1)-state clock model shares the 
three-dimensional XY universality class.

The estimates of $D^*$ and $\beta_c$ obtained from $U_4$ and $U_6$ are
the same up to the digits given here. In contrast, the differences with 
the estimates obtained from $Z_a/Z_p$ and $\xi_{2nd}/L$ are of similar 
size as the statistical errors. These differences are likely due 
to subleading corrections that are not taken into account in the ans\"atze.
We find that the error of $D^*$ obtained from $Z_a/Z_p$ or $\xi_{2nd}/L$
is larger than that of $D^*$ obtained from $U_4$ or $U_6$. Below in section
\ref{Dstar} we give our final estimate of $D^*$.
In the last row of table \ref{betac1} we give our final estimates of 
$\beta_c$, which are mainly based on the analysis of $Z_a/Z_p$ and $\xi_{2nd}/L$. 
The error bars are chosen such that the estimates obtained from 
$Z_a/Z_p$ and $\xi_{2nd}/L$, including their error bars are covered. 
For the inverse critical temperature at the remaining values of $D$
see Appendix \ref{appendixA}.

\subsection{Corrections to scaling}
\label{corrections}
In this section we focus on corrections to scaling. To this end it 
is useful to consider the cumulants $U_4$ and $U_6$ at a fixed value 
of $Z_a/Z_p$ or $\xi_{2nd}/L$ \cite{KlausStefano}.
In particular we take
$Z_a/Z_p=0.32037$ and $\xi_{2nd}/L=0.59238$, which are our estimates
of the fixed point values of these quantities. This means that 
$U_4$ and $U_6$ are taken at $\beta_f$, where $\beta_f$ is chosen such that
either $Z_a/Z_p=0.32037$ or $\xi_{2nd}/L=0.59238$. In the following 
we denote a cumulant at a fixed value of  $Z_a/Z_p$ or $\xi_{2nd}/L$
by $\bar{U}$. Taylor expanding eq.~(\ref{RRfix})  we get
\begin{eqnarray}
 \bar{U} &=&  \bar{U}^* + b(D) L^{-\omega} + c b^2(D) L^{-2 \omega} + ...
                + d(D) L^{-\omega_2} + ... \; \\
          & &     + [f (R_{f}-R^*) + g (D-D^*)] L^{-1/\nu -\omega} + 
               ...  \; ,
\end{eqnarray}
where $R$ denotes either $Z_a/Z_p$ or $\xi_{2nd}/L$.
Note that here $f$ and $g$  are coefficients and not functions.

\begin{figure}
\begin{center}
\includegraphics[width=14.5cm]{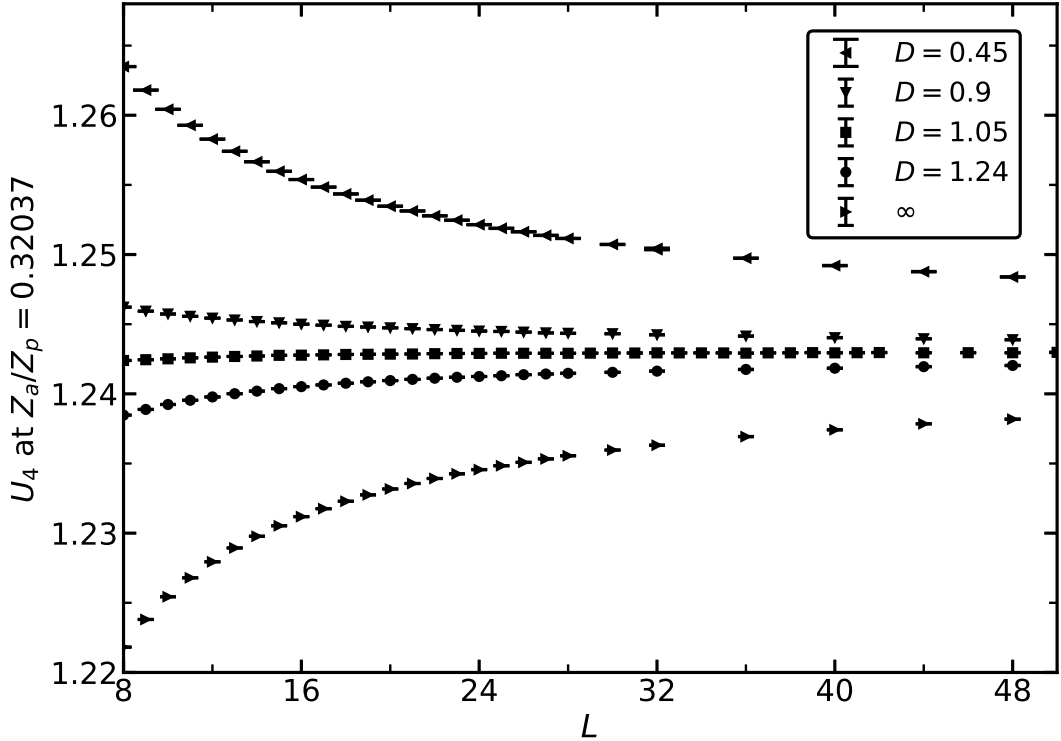}
\caption{\label{Ubar1}
We plot $U_4$ at $Z_a/Z_p=0.32037$ for $N=8$ at $D=0.45$, $0.9$, $1.05$, 
$1.24$, and $\infty$ as a function of the linear lattice size $L$.
}
\end{center}
\end{figure}

\begin{figure}
\begin{center}
\includegraphics[width=14.5cm]{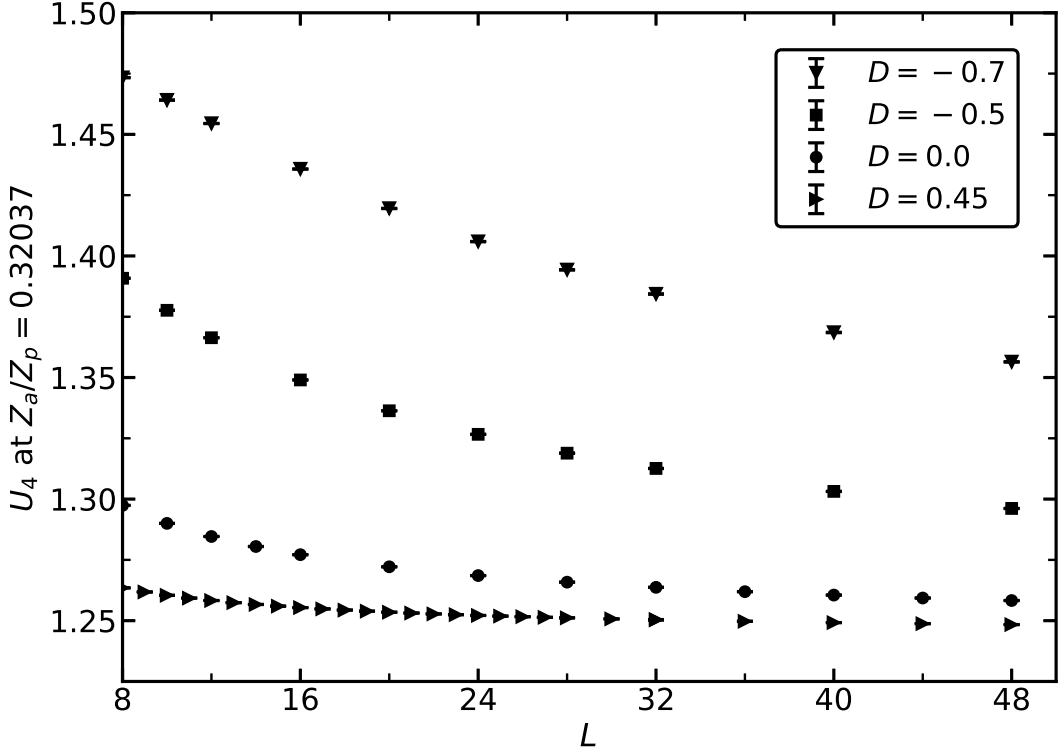}
\caption{\label{Ubar2}
We plot $U_4$ at $Z_a/Z_p=0.32037$ for $N=8$ at $D=0.45$, $0.0$, $-0.5$,
and $-0.7$ as a function of the linear lattice size $L$.
}
\end{center}
\end{figure}

In figure \ref{Ubar1}, as a first step of the analysis, we plot 
$U_4$ at $Z_a/Z_p=0.32037$ for $D=0.45$, $0.9$, $1.05$, $1.24$ and 
$\infty$. We have omitted $D=1.02$ and $1.07$ to keep the figure readable.
For $D=1.05$ we see very little dependence of $\bar{U}_4$ on $L$, 
which confirms that $D=1.05$ is close to $D^*$. For $D=\infty$ we find 
that $\bar{U}_4$ is increasing with increasing lattice size. It is 
approaching the curve for $D=1.05$. For $D=0.45$ we see that $\bar{U}_4$ 
is decreasing and the amplitude of the corrections is roughly equal to that
at $D=\infty$, but with the opposite sign.
Next in figure \ref{Ubar2} we plot $U_4$ at $Z_a/Z_p=0.32037$ for $D=-0.7$,
$-0.5$, $0$, and $0.45$. Going to smaller values of $D$, much larger amplitudes
of the leading correction can be obtained than for $D \rightarrow \infty$. 
Still for $D=-0.7$, where the amplitude of the corrections is the largest, 
the fixed point value is approached as the lattice size increases. This 
indicates that $D=-0.7$ is on the line of second order phase 
transitions. Below we shall study the tricritical point, which is located at 
a smaller value of $D$. 

In the following we determine the exponent of the leading corrections $\omega$ 
and $D^*$, the value of $D$, where the amplitudes of leading corrections 
vanish.

\subsubsection{The correction exponent $\omega$}
\label{Ocorrections}
We performed joints fits of our data for $D=-0.7$, $-0.5$, $0.0$, $0.45$,
$0.9$, $1.02$, $1.05$, $1.07$, $1.24$, and $\infty$. We used the
ansatz
\begin{equation}
\label{manyDfit}
\bar{U} = \bar{U}^* + \sum_{i=1}^{i_{max}} c_i [b(D) L^{-\omega}]^i  + d L^{-\epsilon}.
\end{equation}
In order to avoid ambiguity , we set $c_1=1$. 
In most of our fits we used $\epsilon=2$. 
Furthermore, it is assumed that
$d$ does not depend on $D$. At least for corrections due to the breaking
of the rotational invariance this should be a good approximation.
As a check, we also performed fits without the term $d L^{-\epsilon}$.
Since our final results are taken from fits with $L_{min} \ge 16$, 
the term $d L^{-\epsilon}$ has only a small effect.  
The free parameters of our fits are $\bar{U}^*$, $b(D)$, $c_i$, $\omega$,
and $d$. 

First we fitted all data for all values of $D$ listed above that satisfy
$L \ge L_{min}$.  Here we performed fits with $i_{max} = 2,3,4,5,6$. 
It turns out that the results for $\bar{U}_4^*$, $\bar{U}_6^*$, and 
$\omega$ depend on $i_{max}$.  Let us focus the discussion on $\omega$, which 
is the most important quantity.

\begin{figure}
\begin{center}
\includegraphics[width=14.5cm]{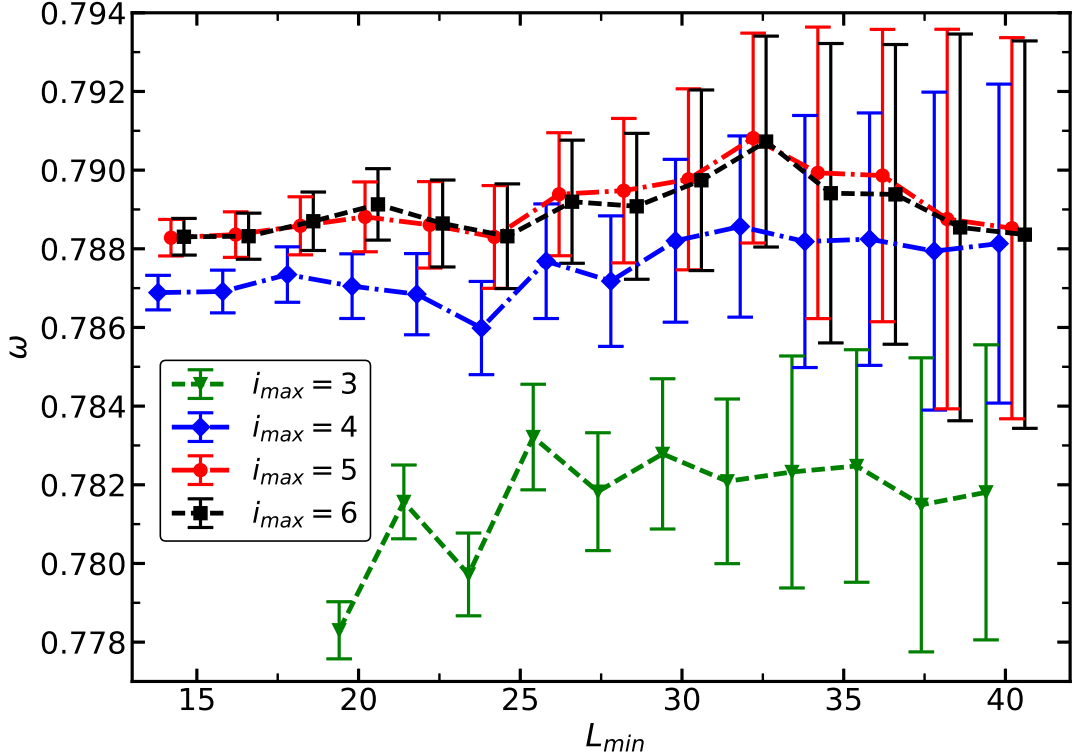}
\caption{\label{omega10}
We plot the estimates of the correction exponent $\omega$ obtained by fitting
$U_4$ at $Z_a/Z_p=0.32037$ using the ansatz~(\ref{manyDfit}), where all 
linear lattice sizes with $L_{min} \le L$ are included.
Data for $N=8$ at $D=-0.7$, $-0.5$, $0.0$, $0.45$,
$0.9$, $1.02$, $1.05$, $1.07$, $1.24$, and $\infty$ are taken into account.
The lines connecting the data points should only guide the eye.
The $L_{min}$ are slightly shifted for different fits to make the 
figure readable.
}
\end{center}
\end{figure}

In figure \ref{omega10} we plot the results obtained from fits with 
$i_{max} = 3,4,5,$ and $6$ of $U_4$ at $Z_a/Z_p=0.32037$. We see that the 
estimates
of $\omega$ are increasing with increasing $i_{max}$. For $i_{max} =5$ and 
$6$ the values saturate. In the plot we give only results that correspond 
to $\chi^2/$d.o.f. $<4$. With increasing $L_{min}$ the $\chi^2/$d.o.f. rapidly converge to  $\chi^2/$d.o.f. $\approx 1$.
As our intermediate result of this set of fits, we take $\omega=0.7886(11)$ from
$i_{max}=5$ and $6$ at $L_{min}=22$.  Performing a similar analysis for 
$U_6$ at $Z_a/Z_p=0.32037$, we arrive at $\omega=0.7880(11)$.

As a check, we have repeated the analysis including fewer values of $D$:
$D=0.45$, $0.9$, $1.02$, $1.05$, $1.07$, $1.24$, and $\infty$. 
Note that for $D=0.45$ the 
amplitude of leading corrections to scaling is, up to the sign, roughly the 
same as for $D=\infty$.  Since we have skipped the data with a large amplitude
of corrections to scaling, already fits with  $i_{max}=2$ are consistent with 
fits using $i_{max}=3$.  As intermediate results we quote $\omega=0.7896(8)$
for $U_4$ and $L_{min}=18$ and $\omega=0.7886(8)$ for $U_6$ and $L_{min}=18$.

Next we analyzed $U_4$ and $U_6$ at $\xi_{2nd}/L=0.59238$.
Our intermediate results for $\omega$ are slightly smaller than those 
obtained above. Furthermore we see a stronger dependence of the 
results on $L_{min}$.

Taking all 10 values of $D$ and $L_{min}=26$ we get 
$\omega=0.7870(14)$ for $U_4$ and 
0.7862(14)  for $U_6$ as intermediate result.
Using only $D \ge 0.45$ we get $\omega=0.7883(21)$ for $L_{min}=30$  
from $U_4$ and $i_{max}=2$.
Based on $U_6$ we arrive at $\omega=0.7875(20)$.

As our final value we quote
\begin{equation}
\label{omegares}
\omega=0.789(4) \;\;.
\end{equation}
The central value is mainly given by the results obtain from $U_4$ and 
$U_6$ at $Z_a/Z_p=0.32037$, since here the estimates depend less on $L_{min}$
than it is the case for fixing $\xi_{2nd}/L=0.59238$.  The error bar is 
chosen such that also the intermediate results obtain for fixing 
$\xi_{2nd}/L=0.59238$ are covered.

\subsubsection{Locating $D^*$}
\label{Dstar}
Next we estimate the value $D^*$ of $D$, where leading corrections
to scaling vanish.  
To this end, we focus again on the neighborhood of $D^*$ and include 
only data for $D=1.02$, $1.05$, and $1.07$ into the analysis.
Since the values of $b(D)$ are small, we have
omitted terms with $L^{-n \omega}$ and $n \ge 2$.  
We made no attempt to discriminate the terms $L^{2-\eta} $ and 
$L^{-\omega_{NR}}$ in our fits. Hence we used a single term with an exponent 
$\epsilon_2 \approx 2$. 
We used the ans\"atze
\begin{eqnarray}
  \label{yyy1}
 \bar{U}(L,D) &=& \bar{U}^*  + b(D) L^{-\epsilon_1} \;\;, \\
  \label{yyy2}
 \bar{U}(L,D) &=& \bar{U}^*  + b(D) L^{-\epsilon_1} + c(D) L^{-\epsilon_2}  \;\;, \\
  \label{yyy3}
 \bar{U}(L,D) &=& \bar{U}^* + b(D) L^{-\epsilon_1} + c(D) L^{-\epsilon_2} + d(D) L^{-\epsilon_3} \;\;.
\end{eqnarray}
Since the values of $D$ differ little, we performed fits where $c$ and $d$ are
the same for all values of $D$. Furthermore $b(D) = b' (D-D^*)$, where 
$b'$ and $D^*$ are the free parameters. 

First we analyzed $U_4$ at $Z_a/Z_p=0.32037$.  
We performed fits without subleading
corrections, with one subleading correction term and with two 
subleading correction terms.   In the case of one subleading correction
term we used the two choices $\epsilon_2 =1.962$ and $\epsilon_2 =2.02$

\begin{figure}
\begin{center}
\includegraphics[width=14.5cm]{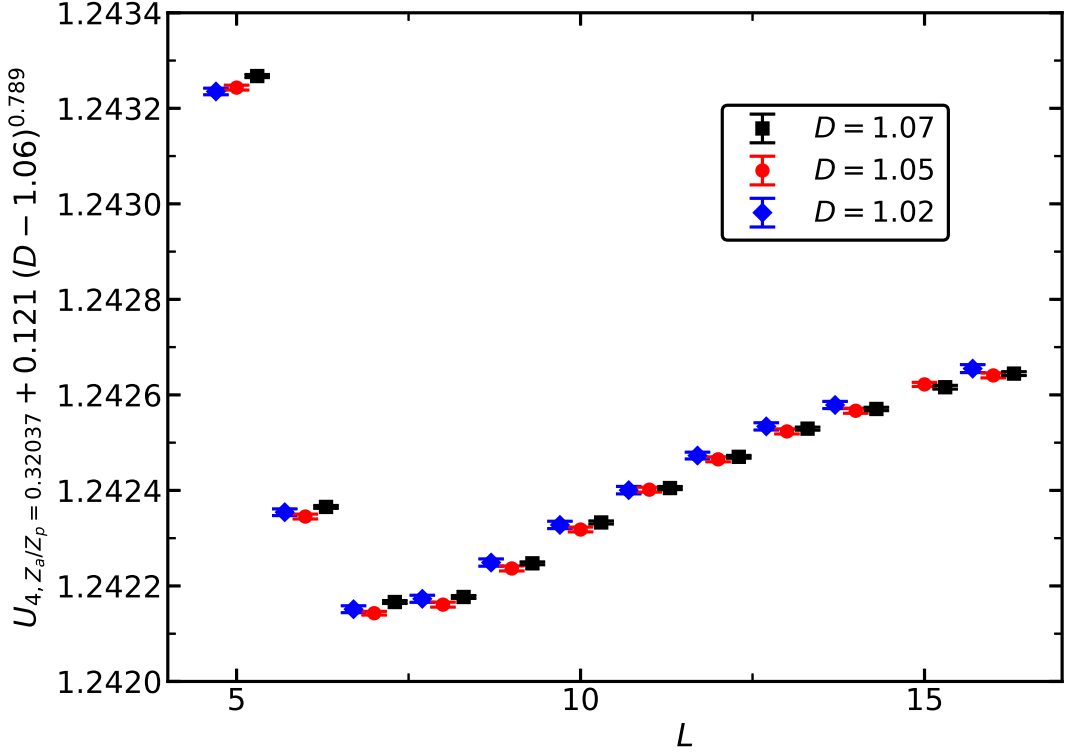}
\caption{\label{Ucoll}
We plot $U_{4}(Z_a/Z_p=0.32037)+ 0.121 (D-1.06) L^{-0.789}$ for $N=8$ at
$D=1.02$, $1.05$, and $1.07$. Note that we have shifted the values of $L$
for $D=1.02$ and $1.07$ to make the figure readable.
}
\end{center}
\end{figure}

Our estimate of the parameter $b'$ for $U_4$ at $Z_a/Z_p=0.32037$ and 
$\epsilon_1=\omega=0.789$ fixed is $b'=-0.121(5)$.  In figure \ref{Ucoll} we plot
$\bar{U}_4 + 0.121 (D-1.06) L^{-0.789}$. 
We find that the data for $D=1.02$, $1.05$, and $1.07$ nicely collapse.
This fact shows that our approximations of $b$, $c$, and $d$ are 
adequate.

\begin{figure}
\begin{center}
\includegraphics[width=14.5cm]{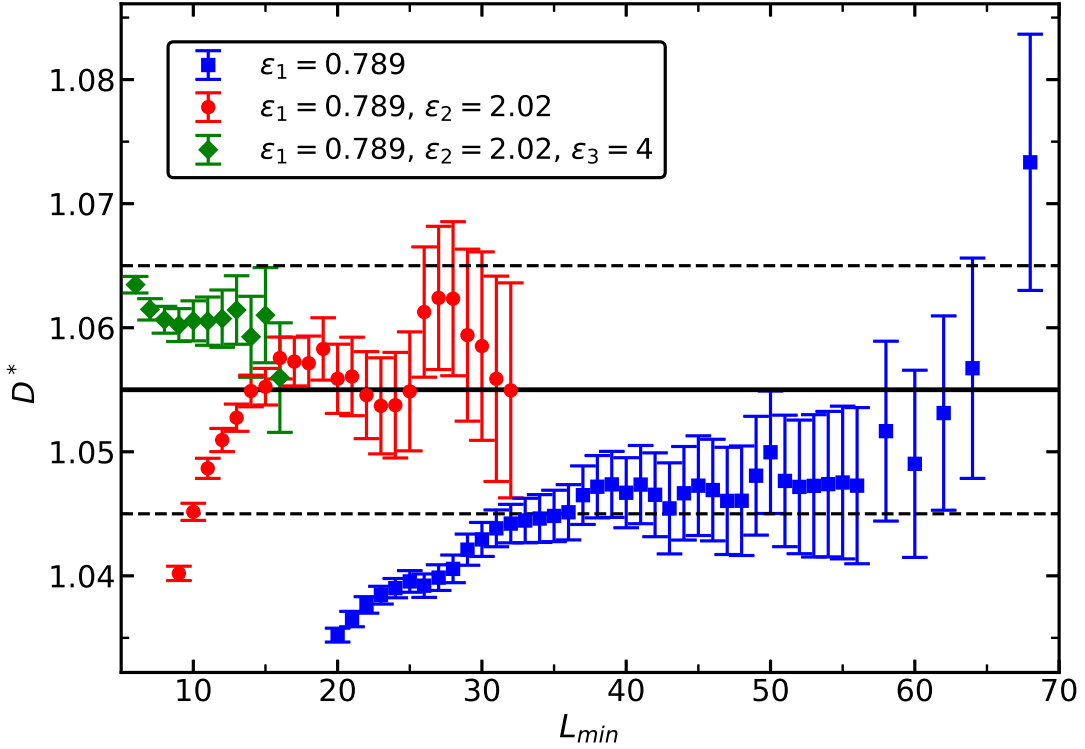}
\caption{\label{DSplot}
We plot estimates of $D^*$ obtained from fits of $U_4$ at $Z_a/Z_p=0.32037$
for $N=8$ at $D=1.02$, $1.05$, and $1.07$ as a function 
of the minimal lattice size $L_{min}$ taken into account.  
The ans\"atze~(\ref{yyy1},\ref{yyy2},\ref{yyy3})
are used. The corresponding correction
exponents are given in the legend. Our preliminary estimate $D^*=1.055(10)$ is 
indicated by the straight solid line. The dashed lines give the error bar.
}
\end{center}
\end{figure}

In figure \ref{DSplot} we plot estimates of $D^*$ obtained by fitting 
$U_4$ at $Z_a/Z_p=0.32037$
with the ans\"atze~(\ref{yyy1},\ref{yyy2},\ref{yyy3}).

Analyzing $U_4$ at $\xi_ {2nd}/L=0.59238$ we get a very similar result.
Overall, the estimates of $D^*$ are shifted by about $0.005$ compared 
with $Z_a/Z_p=0.32037$.  As our final estimate we quote 
\begin{equation}
\label{DSTAR}
 D^* = 1.058(13) 
\end{equation}
that covers both the preliminary estimates obtained from fixing 
$Z_a/Z_p=0.32037$ and $\xi_ {2nd}/L=0.59238$. For a discussion of the 
dependence of $D^*$ on $N$ see appendix \ref{DsonN}.

\subsubsection{The tricritical point}
\begin{figure}
\begin{center}
\includegraphics[width=14.5cm]{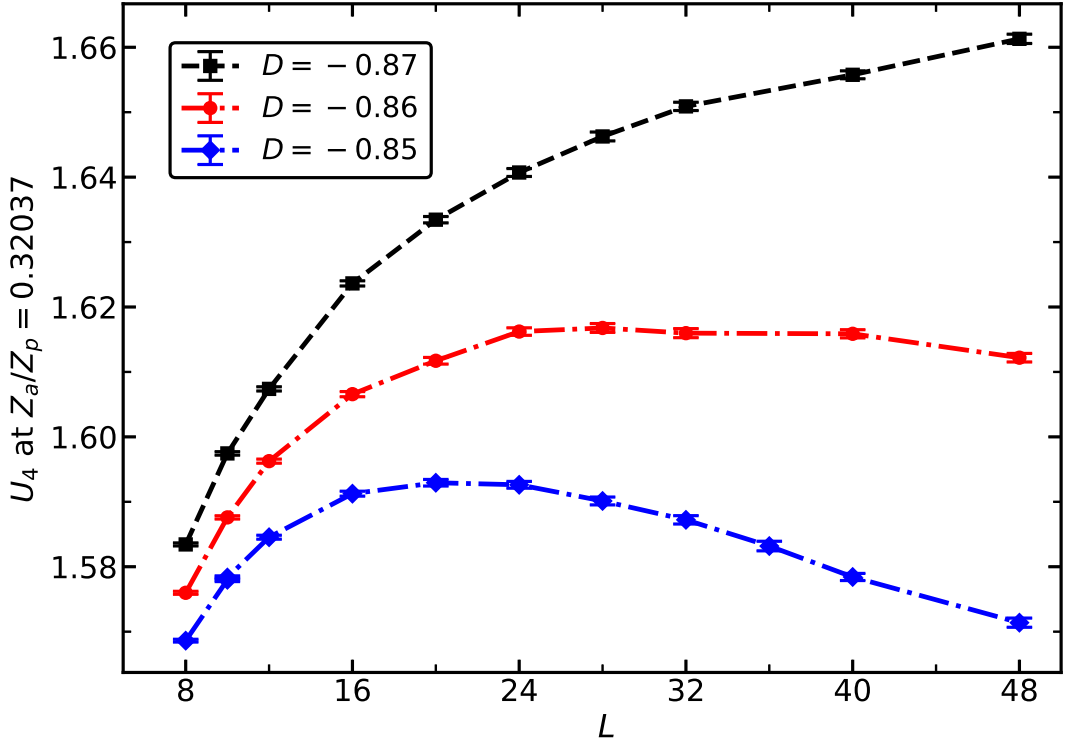}
\caption{\label{U4tri} 
We plot the Binder cumulant $U_4$ at $Z_a/Z_p=0.32037$ for $N=8$ at $D=-0.85$, $-0.86$, and $-0.87$ for 
linear lattice sizes $8 \le L \le 48$. The lines connecting the data points 
should only guide the eye.
}
\end{center}
\end{figure}

The model undergoes a first order phase transition for $D<D_{tri}$.  
We performed  preliminary simulations for a number of $D<D^*$ to roughly
locate $D_{tri}$.  
In figure \ref{U4tri} we plot the Binder cumulant $U_4$ at $Z_a/Z_p=0.32037$ 
for $D=-0.85$, $-0.86$, and $-0.87$, which are close to our preliminary 
estimate of $D_{tri}$.
For $D=-0.87$, the  Binder cumulant is increasing with increasing lattice size 
for the lattice sizes studied. It seems plausible that this behavior 
extends to larger lattice sizes.
In contrast, for $D=-0.86$, and more clearly for $-0.85$, the Binder cumulant 
increases for small lattice sizes, while it decreases for larger ones.
We conclude that $-0.87 < D_{tri} < -0.86$. 

\subsection{The critical exponent $\nu$}
\label{sectnu}

We compute the exponent $\nu=1/y_t$ from the derivative of 
a dimensionless quantity $R_j$ with respect to $\beta$ at a fixed  
value of a second quantity $R_i$, where $R_j$ and $R_i$ might be the same. 
Following the discussion of section \ref{FSStheory} these slopes behave as
\begin{equation}
\label{slope}
\bar{S}_{R,i,j} =
\left . \frac{\partial R_j}{\partial \beta } \right |_{R_i=R_{i,f}} = a 
L^{y_t } \; \left [1 + b L^{-\omega} + ... + c L^{-\omega_{NR}} + ... \right] \;.
\end{equation}
We construct improved slopes by multiplying $\bar{S}_{R,i,j}$ with a certain power $p$
of the Binder cumulant $\bar{U}_4$:
\begin{equation}
\label{improvep}
 \bar{S}_{R,imp} = \bar{S}_R \bar{U}_4^p \;,
\end{equation}
where both $\bar{S}_R$ and $\bar{U}_4$ are taken at $R_{i,f}$.  The exponent $p$ is chosen
such that, at the level of our numerical accuracy, leading corrections to 
scaling are eliminated. This idea is discussed systematically in 
ref. \cite{ourdilute}. To determine $p$, we consider the pairs
$(D_1,D_2)=(0.9,1.24)$ and $(0.45, \infty)$. These pairs are chosen such 
that the amplitude of leading corrections has roughly the same modulus, but
opposite sign.
We fit ratios of $\bar{S}_{R,i,j}$ and $\bar{U}_4$ with the ans\"atze
\begin{equation}
\label{RSfit}
\frac{\bar{S}_{R,i,j} (D_1)}{\bar{S}_{R,i,j}(D_2)} = a_S (1+ b_S L^{- \epsilon_1} )
\end{equation}
and
\begin{equation}
\label{RUfit}
\frac{\bar{U}_4(D_1)}{\bar{U}_4(D_2)} = 1 + b_U L^{- \epsilon_1} \;,
\end{equation}
where we fixed $\epsilon_1=0.789$.   
The exponent $p$ is given by
\begin{equation}
 p = -\frac{b_S}{b_U} \;.
\end{equation}
In table \ref{pimprove} we give our final results for $p$.  These  
are taken from fits for $(D_1,D_2)=(0.9,1.24)$ and $L_{min}=18$.
The statistical error is dominated by eq.~(\ref{RSfit}). 
 In table \ref{pimprove} we give the statistical error only. 
Our numerical results obtained for 
 $(D_1,D_2)=(0.45,\infty)$ are consistent. In the case of 
$(D_1,D_2)=(0.45,\infty)$ we also used fits with one additional
correction term. Note that the results for the
exponent $p$ change very little when we vary $\epsilon_1$ within the 
error bars of eq.~(\ref{omegares}).

\begin{table}
\caption{\sl \label{pimprove}
Numerical result for the exponents $p$ that eliminate 
leading corrections to scaling in $S_R$, eq.~(\ref{improvep}).
}
\begin{center}
\begin{tabular}{ccccc}
\hline
Fixing $\backslash$  Slope of & \phantom{000} $Z_a/Z_p$  \phantom{000} & 
 \phantom{000} $\xi_{2nd}/L$  \phantom{000}  &
 \phantom{0000} $U_4$  \phantom{0000} &  \phantom{0000}$U_6$  \phantom{0000}\\
\hline
$Z_a/Z_p=0.32037$:   & 0.95(3)& 0.30(4) &-2.22(7)&-3.74(7)  \\
$\xi_{2nd}/L=0.59398$:  & 0.60(4) & 0.41(4)  & -2.36(6)  & -3.86(6) \\
\hline
\end{tabular}
\end{center}
\end{table}
As a check, we have computed the RG-exponent $y_t$ for 
$D=\infty$ using the ansatz $\bar{S}_R = a L^{y_t} \; (1 + c L^{-2}) $.
Taking the data for $\bar{S}_{R,imp}$ we get estimates that are consistent with 
our final result obtained below. In contrast, fitting $\bar{S}_R$ without 
improvement, the results differ clearly and depend on the dimensionless
ratio $R$ that is considered.

\subsubsection{Statistical errors}
In the case of the slopes $S_R$ we find a moderate reduction of 
the statistical error when computed at $Z_a/Z_p=0.32037$ or 
$\xi_{2nd}/L=0.59238$ instead of $\beta \approx \beta_c$.
It is of the order of a few percent. In contrast, for the magnetic 
susceptibility that we discuss below, we find a significant reduction.
The relative statistical error of the slope of $Z_a/Z_p$ and
$\xi_{2nd}/L$ is roughly the same. For $U_4$ and $U_6$ for $L=32$ it is 
about twice as large as for $Z_a/Z_p$ and $\xi_{2nd}/L$. With increasing 
lattice size this ratio is shrinking. For $L=512$ roughly a factor of 
$1.8$ remains.  In general, there is a degradation with increasing 
lattice size.  For example, the product of statistics times the square of the 
relative statistical error increases for the slope of $\xi_{2nd}/L$ 
by a factor of $2.4$ going from $L=32$ to $512$. Since we performed a
binning of the data during the simulation, we can not disentangle whether 
this is due to an increasing autocorrelation time or an increasing variance.

\subsubsection{Our final estimate of $y_t$}
The idea of using improved derivatives at  $D \approx D^*$ is
that leading corrections are highly suppressed  and they can be ignored 
safely. In order to obtain our final estimate of $\nu$ we perform 
joint fits of our data obtained for $D=1.05$ and $D=1.07$. 
We use the ans\"atze 
\begin{eqnarray}
\label{slope1}
\bar{S}_R &=& a(D) L^{y_t} \;, \\
\label{slope2}
\bar{S}_R &=& a(D) L^{y_t} (1 + c L^{-\epsilon_1} ) \; ,
\end{eqnarray}
where $\epsilon_1 \approx 2$. This choice is motivated by
the fact that we expect corrections with the exponents $2 -\eta$, 
$\omega_R \approx 2.02$, and $-y_t + \omega \approx 2.278$ and larger ones.
Our final estimates are based on fits with a single correction exponent.

\begin{figure}
\begin{center}
\includegraphics[width=14.5cm]{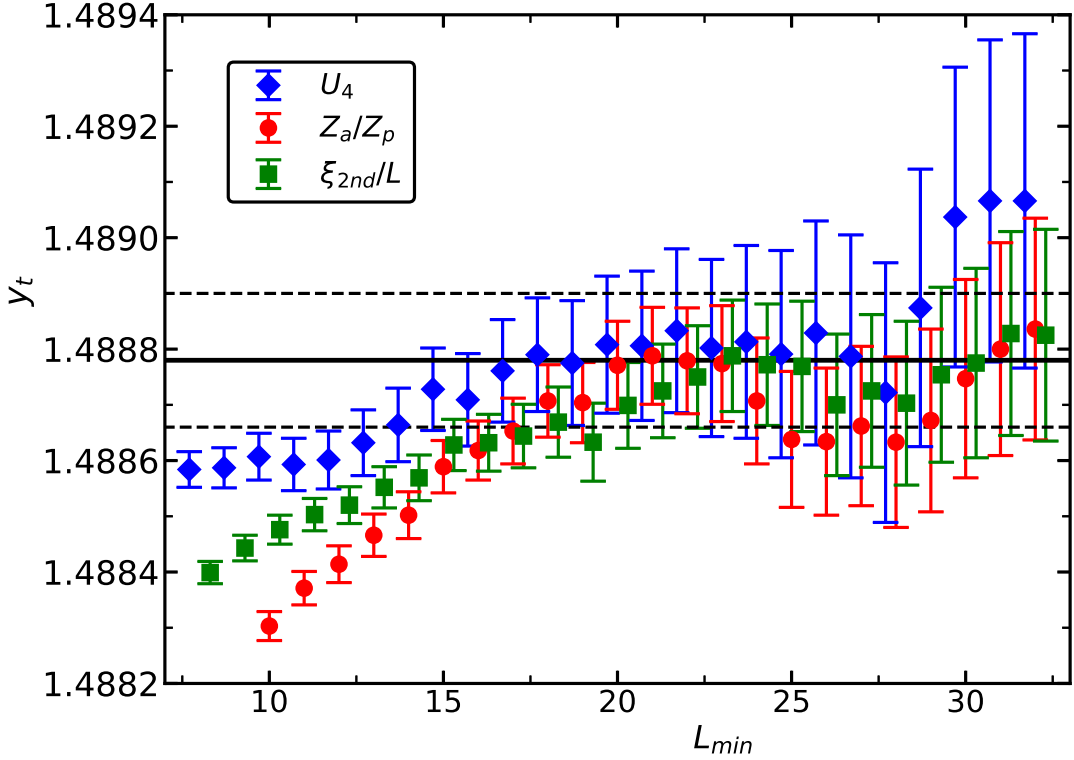}
\caption{\label{fixxic2}
Estimates of the RG-exponent $y_t$ obtained from fitting the improved 
slopes of $U_4$, $Z_a/Z_p$, and $\xi_{2nd}/L$ at 
$\xi_{2nd}/L =0.59238$ for $N=8$ at $D=1.05$ and $1.07$ as a function 
of the minimal linear lattice size $L_{min}$ that is taken into account.
The ansatz~(\ref{slope2}) is used.
To make the figure readable we shifted the values of 
$L_{min}$ by $-0.3$ and $0.3$, for two of the fits. The straight solid 
line gives
our preliminary estimate obtained from the improved slopes at 
$\xi_{2nd}/L =0.59238$. The dashed lines indicate our preliminary error 
estimate.
}
\end{center}
\end{figure}

In figure \ref{fixxic2} we give the results of such fits for 
fixing $\xi_{2nd}/L =0.59238$. The results obtained from the slope of 
$U_6$ are not plotted, since they are similar to those of $U_4$. 
For $Z_a/Z_p$ we get $\chi^2/$d.o.f $=0.871$ 
with $L_{min}=15$. For $\xi_{2nd}/L$ we get $\chi^2/$d.o.f $=1.000$
with $L_{min}=20$. For $U_4$ we get $\chi^2/$d.o.f $=0.815$
already for $L_{min}=7$.
The estimates of $y_t$ obtained from the improved slopes of the 
three different quantities are
consistent starting from $L_{min} \approx 18$. Furthermore the estimates
are increasing with increasing $L_{min}$ up to about $L_{min} =23$. For
$L_{min} =23$, from the slopes of $Z_a/Z_p$ and $\xi_{2nd}/L$ we
read off our preliminary result $y_t = 1.48878(12)$.

\begin{figure}
\begin{center}
\includegraphics[width=14.5cm]{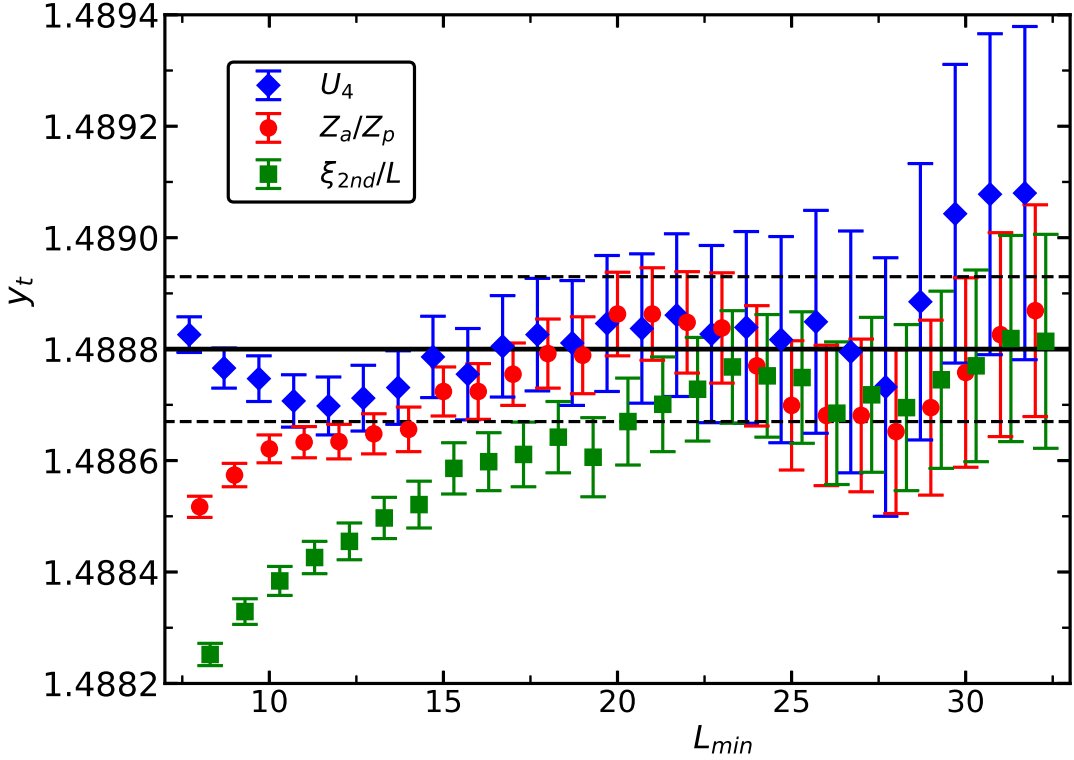}
\caption{\label{fixzc2}
Estimates of the RG-exponent $y_t$ obtained from fitting the improved 
slopes of $U_4$, $Z_a/Z_p$, and $\xi_{2nd}/L$ at 
$Z_a/Z_p =0.32037$ for $N=8$ at $D=1.05$ and $1.07$
as a function 
of the minimal linear lattice size $L_{min}$ that is taken into account.
The ansatz~(\ref{slope2}) is used.
To make the figure readable we shifted the values of 
$L_{min}$ by $-0.3$ and $0.3$, for two of the fits. The straight lines
indicate our preliminary result and its error estimate.
}
\end{center}
\end{figure}

In figure \ref{fixzc2} we give the results of such fits for
fixing $Z_a/Z_p=0.32037$.  In the case of $\xi_{2nd}/L$ we get
$\chi^2/$d.o.f $=1.064$ for $L_{min}=15$.
For $Z_a/Z_p$ we get $\chi^2/$d.o.f $=0.963$ with $L_{min}=10$. In the
case of $U_4$ we get  $\chi^2/$d.o.f $=0.899$ for $L_{min}=8$. Despite this
fact, fully consistent results for $y_t$ among the three quantities 
are only reached for 
$L_{min} \approx 23$. Our preliminary result $y_t=1.48880(13)$ is based 
on the fits of
the slope of $Z_a/Z_p$ and $\xi_{2nd}/L$ for $L_{min} = 23$. In figure
\ref{fixzc2} it is indicated by a straight line. The dashed lines give our 
estimate of the error.

Taking into account both the results from fixing $\xi_{2nd}/L =0.59238$
and  $Z_a/Z_p=0.32037$ we arrive at 
\begin{equation}
\label{finalyt}
 y_t = 1.48879(14)  \,\,.
\end{equation}
The error bar covers  both preliminary estimates, including their
respective error bars. For the critical exponent of the 
correlation length we quote $\nu= 0.67169(7)$.
We repeated the fits using the ansatz~(\ref{slope2}) for
fixing $Z_a/Z_p=0.32$ and $0.321$ and $\xi_{2nd}/L =0.592$ and
$\xi_{2nd}/L =0.593$.  The variation of the results for $y_t$ is well
below the error quoted in eq.~(\ref{finalyt}).

\begin{figure}
\begin{center}
\includegraphics[width=14.5cm]{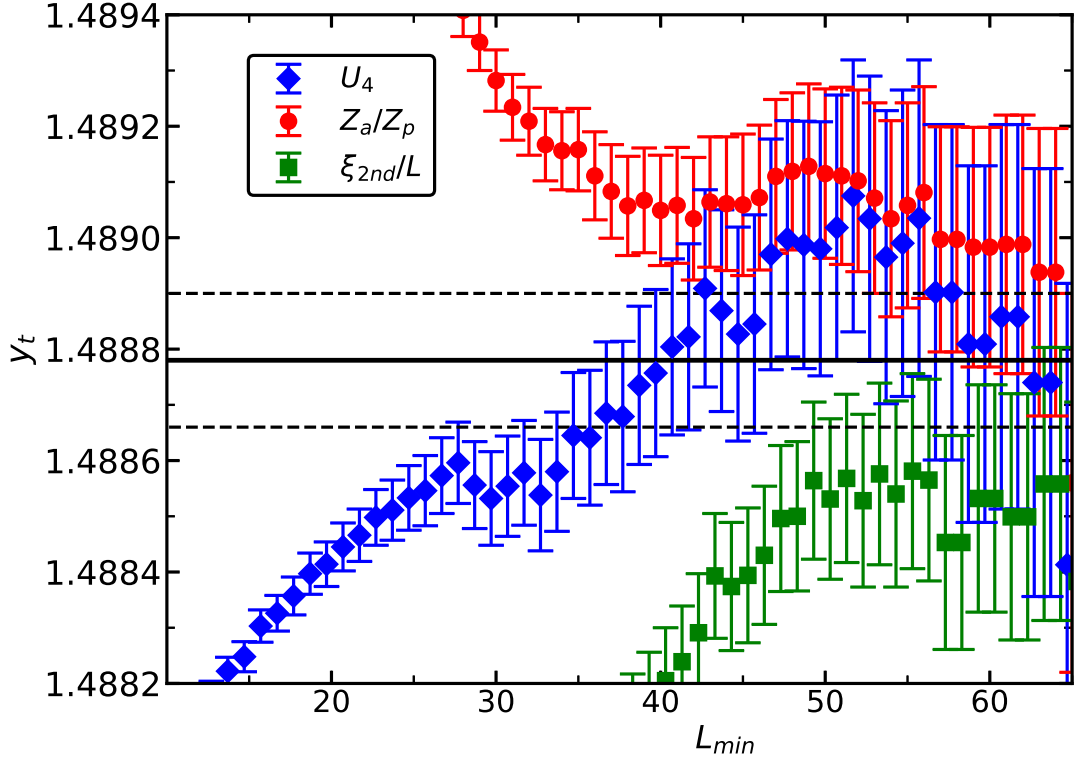}
\caption{\label{fixxic1}
Estimates of the RG-exponent $y_t$ obtained from fitting the improved
slopes of $U_4$, $Z_a/Z_p$, and $\xi_{2nd}/L$ at
$\xi_{2nd}/L=0.59238$ for $N=8$ at $D=1.05$ and $1.07$
as a function of the minimal linear lattice size $L_{min}$ that is
taken into account. 
The ansatz~(\ref{slope1}) is used.
To make the figure readable we shifted the values of
$L_{min}$ by $-0.3$ and $0.3$, for two of the slopes.
}
\end{center}
\end{figure}

Finally, in figure \ref{fixxic1} we show results obtained from fits without 
corrections~(\ref{slope1}). Here we have fixed $\xi_{2nd}/L =0.59238$. 
Fixing $Z_a/Z_p =0.32037$ gives similar results.  We see that
the different estimates of $y_t$ become consistent starting from 
$L_{min} \gtrapprox 60$. As estimate  we read off  $y_t =1.48875(45)$
corresponding to $\nu=0.6717(2)$, which is consistent with the estimate
given above, eq.~(\ref{finalyt}), but less precise.

\subsection{The energy density at the critical point}
\label{sectenergy}
\begin{figure}
\begin{center}
\includegraphics[width=14.5cm]{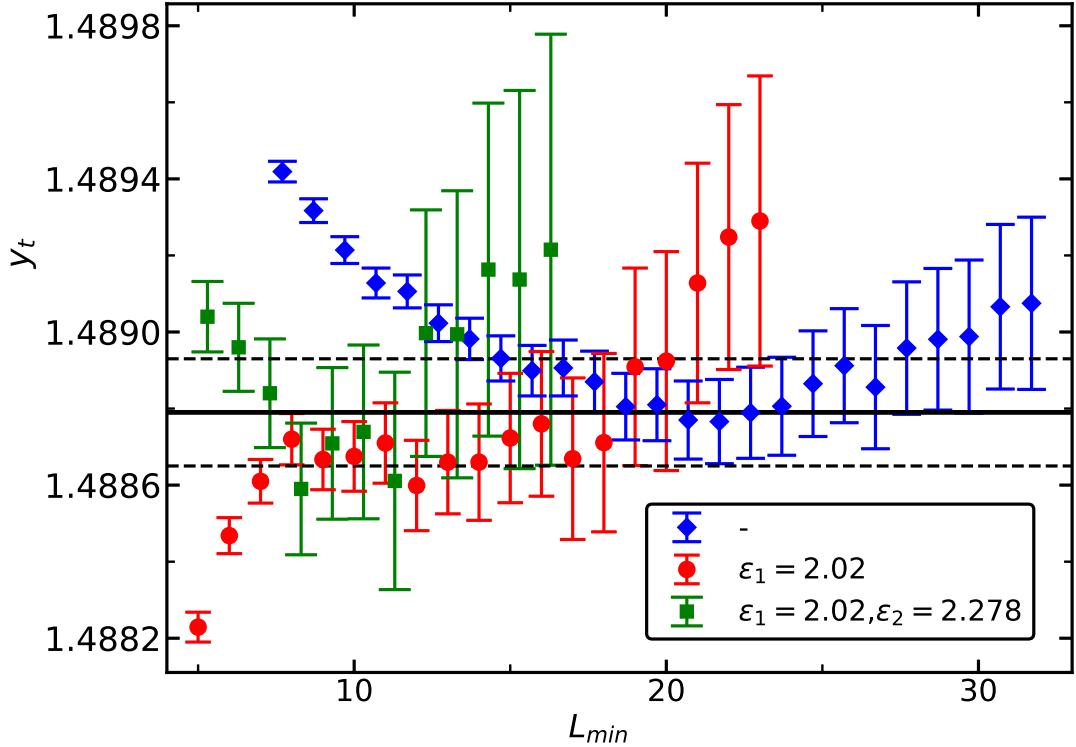}
\caption{\label{enenu}
Estimates for $y_t$ obtained from analyzing the energy density. We fitted 
the data by using the ans\"atze~(\ref{enea1},\ref{enea2},\ref{enea3}). 
The corresponding correction
exponents are given in the legend. $L_{min}$ is the minimal linear lattice
size that is included in the fits.
Data for $N=8$ at $D=1.05$ and $1.07$ are taken into 
account.
For comparison we give the estimate of $y_t$ obtained in the previous 
section by a straight solid line. The dashed lines give the error bar.
}
\end{center}
\end{figure}

We analyzed the energy density, eq.~(\ref{energy}), at our estimates 
of $\beta_c$ for $D=1.05$ and $1.07$. Here we do not consider a fixed
value of $Z_a/Z_p$ or $\xi_{2nd}/L$ since this would generate
contributions $\propto (\beta_f - \beta_c)$ from the analytic
background of the energy density.  Based on eq.~(\ref{enescaling}),
we fitted our data by using the ans\"atze
\begin{eqnarray}
\label{enea1}
E &=& E_0 + a L^{-d +y_t} \;\; ,\\
\label{enea2}
E &=& E_0 + a L^{-d +y_t}  \; \left(1 + c L^{-\epsilon_1} \right)  \;\; , \\
\label{enea3}
E &=& E_0 + a L^{-d +y_t}  \; \left(1 + c L^{-\epsilon_1} + d L^{-\epsilon_2} \right) 
 \;\; ,
\end{eqnarray}
where $\epsilon_1=2.02$ and $\epsilon_2 = y_t + \omega \approx 2.278$.
In our joint fits for $D=1.05$ and $1.07$, $E_0(1.05)$ and $E_0(1.07)$
are both free parameters of the fit. The same holds for $a(1.05)$ and $a(1.07)$.
In contrast, we set $c(1.05)=c(1.07)$ and $d(1.05)=d(1.07)$.
In the case of the ansatz~(\ref{enea1}) we find $\chi^2/$d.o.f.$=0.680$ 
for $L_{min}=15$.  In the case of the ansatz~(\ref{enea2}) we get 
$\chi^2/$d.o.f.$=0.798$  for $L_{min}=8$.  For the
ansatz~(\ref{enea3}) we get  $\chi^2/$d.o.f.$=0.931$ with $L_{min}=5$.
Our results for the RG-exponent $y_t$ are shown in figure \ref{enenu}.
For comparison we give the result obtained in the previous section 
by the solid horizontal line. The estimates of $y_t$ obtained from the 
energy density are consistent with those obtained from the slopes
of dimensionless ratios but a little less precise. Therefore we 
abstain from giving a final estimate of $y_t$ based on the analysis of
this section. 

\subsection{Exponent $\eta$ from the behavior of the magnetic susceptibility $\chi$}
\label{secteta}
As observed in previous work \cite{KlausStefano}, we find that the statistical error 
of $\chi$ is reduced, when computed at a fixed value of a phenomenological
coupling compared with the error at a given value of $\beta \approx \beta_c$.
Comparing $U_4$, $Z_a/Z_p$ and $\xi_{2nd}/L$ we find that the reduction 
is clearly the largest for fixing $\xi_{2nd}/L=0.59238$. For example 
for $D=1.07$ and $L=512$ we find a reduction of the statistical error by a factor 
of about 3.3 compared with $\chi$ at $\beta=0.55888340$. 
This factor is slowly increasing with increasing lattice
size.

Also here we analyzed the improved quantities 
\begin{equation}
\label{chiimp}
\bar{\chi}_{imp} = \bar{\chi}  \bar{U}_4^p \;,
\end{equation}
where both $\chi$ and $U_4$ are taken either at $Z_a/Z_p =0.32037$ or
$\xi_{2nd}/L =0.59238$. We computed the exponent $p$ in a similar way as in the previous
section for $S_R$. Therefore we skip a detailed discussion and only report 
our results:  $p=-0.97(2)$ and $-0.45(1)$ for $Z_a/Z_p =0.32037$ 
and $\xi_{2nd}/L =0.59238$, respectively. 

We fitted our data with the ans\"atze 
\begin{eqnarray}
\label{eta1}
\bar{\chi}_{imp} &=& a L^{2-\eta} \;, \\
\label{eta2}
\bar{\chi}_{imp} &=& a L^{2-\eta} + b \;, \\
\label{eta3}
\bar{\chi}_{imp} &=& a L^{2-\eta}  (1 + c L^{-\epsilon_2})   + b  \;.
\end{eqnarray}
In the case of eq.~(\ref{eta3}), we  fixed either $\epsilon_2=2.02$
or $\epsilon_2=4$.

\begin{figure}
\begin{center}
\includegraphics[width=14.5cm]{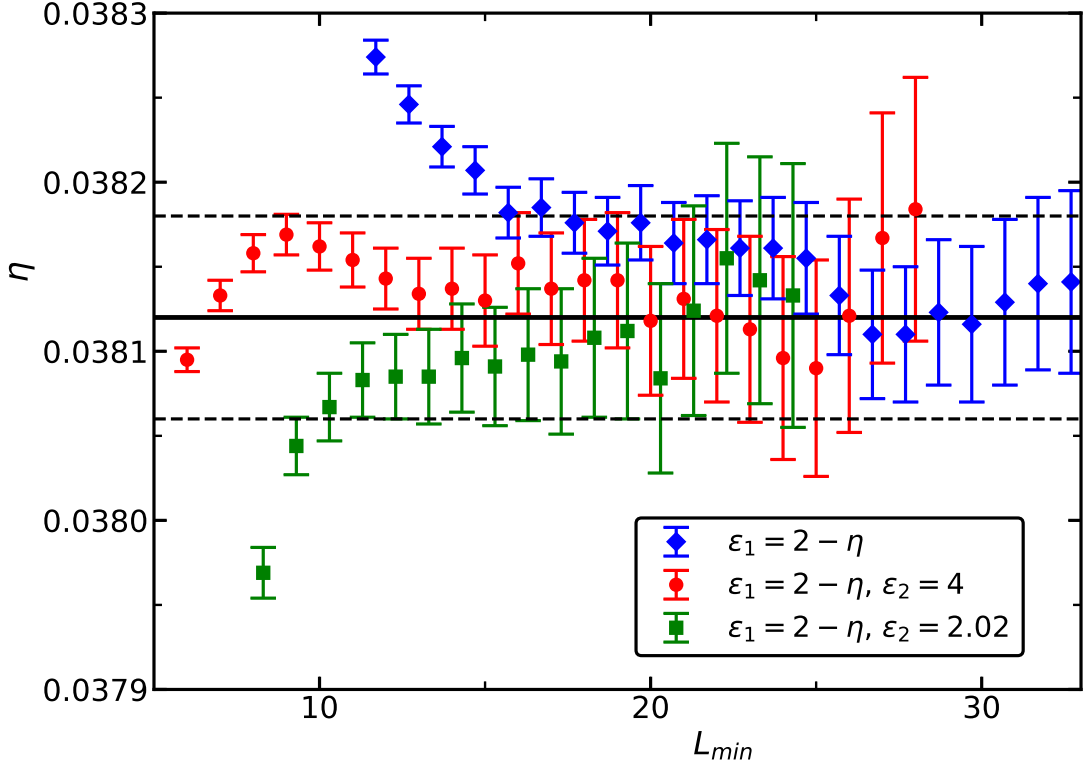}
\caption{\label{eZI}
Estimates of the critical exponent $\eta$ obtained from fitting the
improved magnetic susceptibility $\chi_{imp}$ at $Z_a/Z_p=0.32037$ 
for $D=1.05$ and 
$1.07$ as a function of the minimal linear lattice size $L_{min}$ that is 
taken into account. The ans\"atze~(\ref{eta2},\ref{eta3}) are used.
To make the figure readable we shifted the values of $L_{min}$ by $-0.3$ and $0.3$, for two of the fits.
}
\end{center}
\end{figure}

Let us first discuss the analysis of the data for $Z_a/Z_p =0.32037$ 
fixed. In figure \ref{eZI} we plot our estimates of $\eta$ obtained 
by using the ans\"atze~(\ref{eta2}) and (\ref{eta3}).
In figure \ref{eZI}, the analytic background is indicated by
$\epsilon_1=2-\eta$.  In the case of ansatz~(\ref{eta2})
we find $\chi^2/$d.o.f. $=0.899$ for $L_{min}=16$.  For the 
ansatz~(\ref{eta3}) $\chi^2/$d.o.f. is less than one starting from 
$L_{min}=11$ and $8$ for $\epsilon_2=2.02$ and $\epsilon_2=4$, respectively.
As our preliminary estimate we take $\eta=0.03812(6)$. Fitting without correction term, eq.~(\ref{eta1}),
$\chi^2/$d.o.f. $=0.95$ is reached for  $L_{min}=40$. However the estimates of $\eta$ are 
further increasing with increasing $L_{min}$. For $L_{min}=96$ the estimates seem to level off.
We get $\eta=0.03813(15)$ for $L_{min}=96$.

\begin{figure}
\begin{center}
\includegraphics[width=14.5cm]{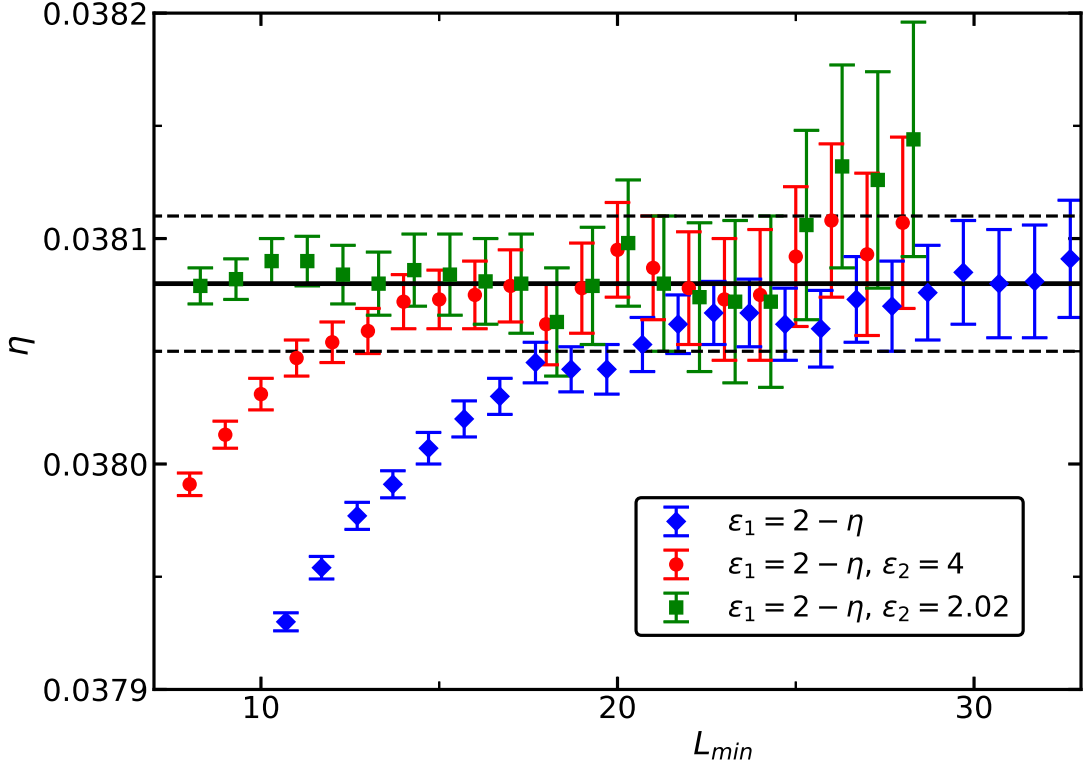}
\caption{\label{eXI}
Estimates of the critical exponent $\eta$ obtained from fitting the 
improved magnetic susceptibility $\chi_{imp}$ at $\xi_{2nd}/L =0.59238$
for $D=1.05$ and 
$1.07$ as a function of the minimal linear lattice size $L_{min}$ that is 
taken into account. The ans\"atze~(\ref{eta2},\ref{eta3}) are used.
To make the figure readable we shifted the values of $L_{min}$ by $-0.3$ and $0.3$, for two of the fits.
}
\end{center}
\end{figure}

Next we turn to $\xi_{2nd}/L =0.59238$. In figure \ref{eXI} we plot our estimates of $\eta$ obtained
by using the ans\"atze~(\ref{eta2}) and (\ref{eta3}). In the case of ansatz~(\ref{eta2})
we find $\chi^2/$d.o.f. $=1.053$ for $L_{min}=18$.  For the
ansatz~(\ref{eta3}) $\chi^2/$d.o.f. is approximately one starting from
$L_{min}=18$ and $14$ for $\epsilon_2=2.02$ and $\epsilon_2=4$, respectively.
As our preliminary estimate we take $\eta=0.03808(3)$. Fitting without correction term, eq.~(\ref{eta1}),
$\chi^2/$d.o.f. $=1.336$ is reached for  $L_{min}=64$.  For $L_{min}=96$ we get $\eta=0.03808(7)$.

We also analyzed the data for $\chi$ without improvement, eq.~(\ref{chiimp}). We do not report the results 
in detail. They are consistent with those reported above.

As our final result we quote
\begin{equation}
\label{finalchi}
 \eta = 0.03810(8) \;,
\end{equation}
which is chosen such that the results obtained by using the 
ans\"atze~(\ref{eta2},\ref{eta3}) for fixing $Z_a/Z_p=0.32037$ and
$\xi_{2nd}/L =0.59238$ are covered. 
As the last check we repeated the fits using the ansatz~(\ref{eta2}) for 
fixing $Z_a/Z_p=0.32$ and $0.321$ and $\xi_{2nd}/L =0.592$ and 
$\xi_{2nd}/L =0.593$.  The variation of the results for $\eta$ is well
below the error quoted in eq.~(\ref{finalchi}).

\section{Summary and conclusions}
We have studied a generalized clock model on the 
simple cubic lattice by using a finite size scaling analysis. In the 
case of the $N$-state clock model, for $N \ge 5$, 
at the critical point, with increasing length scale, the $Z_N$ symmetry is 
enhanced to $O(2)$; See for example \cite{HoSu03}.
In the generalized model, denoted by $(N+1)$-state clock model,
$(0,0)$ is added as allowed value of the spin. The parameter 
$D$, which controls the relative weight of $(0,0)$, can be 
tuned such that the amplitude of leading corrections to scaling 
vanishes.
We were aiming at accurate estimates of critical exponents for the 
three-dimensional XY universality class.
Our motivation to study the $(N+1)$-state clock model is that the 
simulation requires less CPU-time and less memory than  that
of a model with $O(2)$ symmetry at the microscopic level. 
 
In the main part of our study we considered $N=8$. 
The RG-exponent related with a $Z_8$ symmetric perturbation of the $O(2)$
invariant fixed point takes the value $y_{N=8}=-5.278(9)$ \cite{Debasish}. 
Hence deviations from $O(2)$ symmetry vanish rapidly with increasing 
lattice size and can be ignored in the finite size analysis of the data. 
For $N=8$
we find even for critical temperatures, which
depend on the microscopic details of the model, only little differences 
compared with the $N \rightarrow \infty$ limit. 
For a detailed discussion see appendix \ref{Ndependence}.
In total we have spend the equivalent 50 years of CPU time on a single core of
a Intel(R) Xeon(R) CPU E3-1225 v3 running at 3.20GHz.

Simulating the model for a large range of the parameter $D$ we determined
the exponent of the leading correction $\omega=0.789(4)$ accurately. 
We located the tricritical point in the phase diagram. The corresponding 
$-0.87 < D_{tri} < -0.86$ is clearly smaller than $D^*=1.058(13)$, 
where the amplitude of the leading correction vanishes.  
Focusing on the neighborhood of $D^*$ we obtain $\eta = 0.03810(8)$ and
$\nu= 0.67169(7)$, which are consistent with but more accurate than previous 
Monte Carlo results \cite{XY2,Xu19}.
The discrepancy with the experiments on the $\lambda$-transition of $^4$He
\cite{Lipa96,Lipa00,Lipa03} is not dissolved.  Note that the results of ref. \cite{che19}, 
which appeared after we had submitted the first version of this paper, 
nicely agree with ours.

We determined the inverse of the critical temperature $\beta_c$ for 
various values of $D$ accurately. This is important information for 
coming studies. We plan compute two- and three-point functions at criticality 
on large lattices, similar to ref. \cite{myStructure}, in order to get 
estimates for operator product expansion coefficients.

One might also study the low temperature phase of the improved $(N+1)$-state 
clock model.
The consequences of the fact that a $Z_N$ symmetric perturbation of 
the $O(2)$ symmetric fixed point is dangerously irrelevant in the 
low temperature phase are debated
in the literature, as can be seen in \cite{Shao19} and references therein.

\section{Acknowledgement}
This work was supported by the DFG under the grant No HA 3150/5-1.

\appendix

\section{The inverse critical temperature for $N=8$}
\label{appendixA}
Here we compute $\beta_c$ for those values of $D$ that are not
consider in section \ref{betac0}. To this end we analyze the behavior of 
$Z_a/Z_p$ and $\xi_{2nd}/L$.
We fit our data with the ans\"atze
\begin{eqnarray}
 R(\beta_c) &=& R^* + b L^{-\epsilon} \; , \\
 R(\beta_c) &=& R^* + b L^{-\epsilon} + c L^{-2 \epsilon}  \; , \\
 R(\beta_c) &=& R^* + b L^{-\epsilon} + c L^{-2 \epsilon}  + d L^{-3 \epsilon}
\end{eqnarray}
using $\epsilon=0.789$. As in section \ref{betac0}, we compute 
$R(\beta)$ by using its  Taylor expansion around $\beta_s$ up to the 
third order. The free parameters of the fits are 
$\beta_c$, $b$, $c$, and $d$. $R^*$ is fixed by the numerical results
obtained in section \ref{betac0}.  Our results for $Z_a/Z_p =0.32037$ 
are summarized in table \ref{betacmehr}. The results obtained for 
$\xi_{2nd}/L=0.59238$ are compatible.

\begin{table}
\caption{\sl \label{betacmehr}
We give our numerical result for the inverse critical 
temperature $\beta_c$ for $N=8$ at the values of $D$ not considered in 
section \ref{betac0}. Here we use $(Z_a/Z_p)^* =0.32037(6)$. The number
given in $[]$ is the error due to the uncertainty of $(Z_a/Z_p)^*$. 
}
\begin{center}
\begin{tabular}{cl}
\hline
$D$ &   \mc{1}{c}{$\beta_c $} \\
\hline
$\infty$&  0.45416467(10)[7]    \\
1.24   &  0.54365020(30)[10]    \\    
0.9    &  0.57645235(30)[11]    \\  
0.45   &  0.63625739(10)[8]     \\       
0.0    & 0.7191494(3)[1]      \\
-0.5   & 0.8423571(7)[1]      \\
-0.7   & 0.9008977(10)[1]     \\
\hline
\end{tabular}
\end{center}
\end{table}

\section{The $N$-dependence of the inverse critical temperature and $D^*$}
\label{Ndependence}
\subsection{The Caley tree}
In order to get a first idea,  we have computed numerically $\beta_c$
for the model put on a Caley tree with the coordination number $z=6$.
The phase transition is of mean-field type. However $\beta_c$ for the Caley 
tree should be a better approximation of $\beta_c$ for the three-dimensional
model than simple mean-field.

For given values of $D$ and $\beta>\beta_c$ we computed the magnetization. 
Estimates of the inverse critical temperature are obtained by solving 
\begin{equation}
\label{powermean}
 m = c (\beta-\beta_c)^{1/2}
\end{equation}
for two different values of $\beta$ with respect to $c$ and $\beta_c$.
Iteratively we diminish $\beta-\beta_c$ until corrections to eq.~(\ref{powermean}) 
can be ignored. This way we obtain the critical temperature up to about
10 accurate digits.
 
We computed 
$\beta_c$ for $D = \infty$, $1.0$ and $0.0$ and $N=5$, $6, ..., 12$.
Our results are given in table \ref{betaCaley}.
We find that, at the level of our precision, the results
are identical starting from $N=10$ for  $D=\infty$ and $D=1$. For 
$D=0$ this holds starting from $N=11$. Deviations from the limit 
$N \rightarrow \infty$ seem to increase with decreasing $D$. 
The approach $N \rightarrow \infty$ is compatible with an exponential
decay with a large, $D$ dependent, decay rate.

\begin{table}
\caption{\sl \label{betaCaley}
We give our numerical result for the inverse critical temperature
$\beta_{c,Caley}$ for the Caley tree with coordination number $z=6$. 
}
\begin{center}
\begin{tabular}{cccc}
\hline
$N$ $\backslash$  $D$  &   $\infty$     &      1         &       0  \\
\hline             
  5   &     0.4081307306   & 0.5224090169  &  0.6890295689 \\ 
  6   &     0.4082712294   & 0.5227444788  &  0.6898803344 \\
  7   &     0.4082770202   & 0.5227621638  &  0.6899394147 \\
  8   &     0.4082772183   & 0.5227629375  &  0.6899428166 \\
  9   &     0.4082772241   & 0.5227629665  &  0.6899429844 \\
 10   &     0.4082772243   & 0.5227629675  &  0.6899429916 \\
 11   &     0.4082772243   & 0.5227629675  &  0.6899429919 \\
 12   &     0.4082772243   & 0.5227629675  &  0.6899429919 \\
\hline             
\end{tabular}
\end{center}
\end{table}

\subsection{$N$-dependence of $\beta_c$: three-dimensional model}
We performed simulations for $N \ne 8$ for a small number of lattice sizes.
We determined $\beta_{f,Z_a/Z_p=0.32037}$, where $Z_a/Z_p=0.32037$ and
$\beta_{f,\xi_{2nd}/L=0.59238}$, where $\xi_{2nd}/L=0.59238$.  
Since the difference of $\beta_c$ for different values of $N$ 
is essentially related to the microscopic details of the model at small 
scales,
we expect that differences or ratios of $\beta_f$ obtained for 
moderate lattice sizes are 
good approximations of the differences or ratios of $\beta_c$.
Note that $Z_a/Z_p$ is only defined for even values of $N$. 
We study the ratio
\begin{equation}
\label{bratio}
r(L)= \frac{\beta_{f,N=8}(L)}{\beta_{f,N}(L)} \;.
\end{equation}
As discussed in section \ref{FSStheory}, there is an $N$ dependence 
of all scaling fields. 
In particular there should be a, even though
small, dependence of the scaling field related to the leading correction 
to scaling. Therefore we expect that
\begin{equation}
r(L)= \frac{\beta_{c,N=8}}{\beta_{c,N}} + c L^{-\epsilon} + ... \; .
\end{equation}
where $\epsilon=1/\nu+\omega$ is the exponent related with the leading 
correction.
We performed simulations for  $D = \infty$, $1.07$, and $D=1.02$. 
Let us first discuss our results for $D = \infty$.  For $N=6$ we 
simulated the linear lattice sizes $L=32$, $36$, and
$40$. The ratios, eq.~(\ref{bratio}),  for these three lattice sizes are
consistent within their error bars. The average is given in table 
\ref{betaRXY}. For $N=7$ we simulated the lattice sizes $L=36$ and $40$.
For $N=10$ we simulated the lattice sizes $L=32$, $40$ and $48$.
Also for these two values of $N$, the averages are given in 
table \ref{betaRXY}.  In addition we make use of the estimates 
$1/\beta_{c,N=5} = 2.20502(1)$ and $1/\beta_{c,N=6} = 2.20201(1)$ 
reported in \cite{Shao19}.  Note that for $N=6$ the result of \cite{Shao19} is 
fully consistent with ours. Similar to the Caley tree approximation, we
see a rapid convergence of $\beta_{c,N}$ with $N \rightarrow \infty$.
Already for $N=8$ and $10$, we can not find a difference at our 
level of accuracy. Extrapolating the ratios for smaller 
values of $N$ we get $\beta_{c,N=8}/\beta_{c,9} \approx 0.99999985$.
At our level of precision, the same ratio holds for all $N \ge 9$. 
Using this estimate, we arrive at $\beta_{c,XY} = 0.45416474(10)[7]$. 
In table \ref{betacXY} we summarize estimates of $\beta_{c,XY}$
given in the literature.

Next let us discuss the results for $D=1.07$. Here we simulated the 
linear lattice sizes $L=64$ for $N=6$, $L=32$, $40$, $48$, and $64$
for $N=7$ and  $L=48$ and $64$  for $N=12$.  The averages of the 
ratios of $\beta_f$ are reported in table \ref{betaRXY}. Similar to 
$D=\infty$ we see a rapid convergence of $\beta_{c,N}$, which is however
slightly slower than it is the case for $D=\infty$.  In particular 
our estimate for $\beta_{c,N=8}/\beta_{c,N=12}$ differs from $1$  
by about $3.6$ times the error bar. Extrapolating the results for $N<8$
we arrive at $\beta_{c,N=8}/\beta_{c,N>8} \approx 0.9999995$.

Finally for $D=1.02$ we have simulated $L=4$, $5, ...$, $14$, $16$, $18$, $20$,
and $64$ for $N=6$.  These simulations were performed at an early stage of the study, 
mainly to determine the correction exponent $y_6$. 
Here we see a dependence of the ratio $r$,  eq.~(\ref{bratio}), 
on the lattice size $L$. First we analyzed the results obtained for 
$\beta_{f,Z_a/Z_p=0.32037}$. We fitted our data with the ansatz
\begin{equation}
 r(L) = a + c L^{-\epsilon}  \;\;,
\end{equation}
using the numerical value $\epsilon=1/\nu+\omega=2.27779$. 
Including data with $L \ge 8$ we get $a=1.0001772(3)$, $c = 0.00155(12)$
and $\chi^2/$d.o.f.$=0.70$. 
The analysis of the data for $\beta_{f,\xi_{2nd}/L=0.59238}$ gives 
very similar results. Our final estimates are given in table \ref{betaRXY}.

\begin{table}
\caption{\sl \label{betaRXY}
We give our numerical estimates for the ratio $r=\beta_{c,N=8}/\beta_{c,N}$
obtained from $\beta_{f,Z_a/Z_p=0.32037}$ and 
$\beta_{f,\xi_{2nd}/L=0.59238}$. In addition results based on ref. \cite{Shao19}
are reported.
}
\begin{center}
\begin{tabular}{crlll}
\hline
$D$ & \mc{1}{c}{$N$} &\mc{1}{c}{$Z_a/Z_p=0.32037$} & 
  \mc{1}{c}{$\xi_{2nd}/L=0.59238$} & \mc{1}{c}{ref. \cite{Shao19}}\\
\hline
$\infty$ & 5  & \mc{1}{c}{-}    & \mc{1}{c}{-}    & 1.001442(5)   \\  
$\infty$ & 6  & 1.00007847(22)  & 1.00007837(20)  & 1.000075(5)   \\ 
$\infty$ & 7  & \mc{1}{c}{-}    & 1.00000362(21)  & \mc{1}{c}{-} \\
$\infty$ &10  & 1.00000015(21)  & 0.99999984(20)  & \mc{1}{c}{-} \\
\hline
  1.07   & 6  & 1.00017147(32)  & 1.00017141(29)  & \mc{1}{c}{-} \\
  1.07   & 7  & \mc{1}{c}{-}    & 1.00000946(15)   & \mc{1}{c}{-} \\
  1.07   & 12 & 0.99999938(16)  & 0.99999947(16)  & \mc{1}{c}{-} \\
\hline
  1.02   & 6  & 1.0001772(3)    & 1.0001769(3)    & \mc{1}{c}{-}   \\
\hline
\end{tabular}
\end{center}
\end{table}

\begin{table}
\caption{\sl \label{betacXY}
We summarize results from the literature for the inverse critical
temperature of the XY model on the simple cubic lattice.
}
\begin{center}
\begin{tabular}{ccl}
\hline
ref. & year &  \mc{1}{c}{$\beta_c$} \\
\hline
\cite{Deng05} & 2005 & 0.4541655(10) \\ 
\cite{XY2}    & 2006 & 0.4541652(5)[6]\\
\cite{Lan12}  & 2012 & 0.45416313(20) \\ 
\cite{Lan12}  & 2012 & 0.45416742(12) \\ 
\cite{Komura} & 2014 & 0.4541664(12) \\  
\cite{Xu19}   & 2019 & 0.45416466(10) \\    
\hline
this work     & 2019 & 0.45416474(10)[7] \\
\hline
\end{tabular}
\end{center}
\end{table}

\subsection{$N$-dependence of $D^*$}
\label{DsonN}
As discussed in section \ref{FSStheory},
the value of $D^*$ depends on $N$. To get a numerical estimate, we analyze the 
Binder cumulant $\bar{U}_4$ at either $Z_a/Z_p=0.32037$ or $\xi_{2nd}/L=0.59238$
at values of $D$ close to $D^*$.

First we estimate the slope of the correction amplitude close to $D^*$ for $N=8$ by 
fitting the data with the ansatz
\begin{equation}
\bar{U}_4(N=8,D=1.07) - \bar{U}_4(N=8,D=1.02)  = b_d L^{-\omega}  \;,
\end{equation}
where we have fixed $\omega=0.789$, or
\begin{equation}
\bar{U}_4(N=8,D=1.07) - \bar{U}_4(N=8,D=1.02)  = b_d L^{-\omega} + c_d  L^{-2} \;.
\end{equation}
In the following we assume that the dependence of 
\begin{equation}
\left. \frac{\mbox{d} b}{\mbox{d} D} \right|_{D=D^*}  \approx \frac{b_d}{0.05} 
\end{equation}
on $N$ can be ignored.  We get $b_d=-0.00616(10)$ for fixing $Z_a/Z_p=0.32037$  and   
$b_d=-0.00705(16)$  for fixing $\xi_{2nd}/L=0.59238$.

In the second step, we analyze how  $\bar U_4$ changes with $N$
at a fixed value of $D$. To this end we define
\begin{equation}
 \Delta_U(N_1,N_2,D) = \bar{U}_4(N_1,D) - \bar{U}_4(N_2,D) \;,
\end{equation}
where here $N_2=8$. We fitted our data with the ans\"atze
\begin{equation}
\label{simpleD}
 \Delta_U(N_1,N_2,D) = \Delta_b(N_1,N_2,D) L^{-\omega} \;,
\end{equation}
where we have fixed $\omega=0.789$ and 
\begin{equation}
\label{doubleD}
\Delta_U(N_1,N_2,D) = \Delta_b(N_1,N_2,D) L^{-\omega} + \Delta_c(N_1,N_2,D) L^{-\epsilon} \;,
\end{equation}
where we fixed $\epsilon=2$. In the case of $N_1=6$ we used in addition $\epsilon=2.4$. 
The shift in $D^*$ is  given by 
\begin{equation}
\label{shiftD}
D^*(N_1) - D^*(8) \approx - \Delta_b(N_1,8,D) \frac{0.05}{b_d}  \;.
\end{equation} 

For the purpose of this section, we have simulated the linear lattice
size $L=4$, $5$, $6$,..., and $16$ for $N=10$ at $D=1.07$ with a statistics
similar to that for $N=8$. It turns out that $\Delta_U(10,8,1.07)$ is 
compatible with zero for most of the lattice sizes. Fitting the data for
$L \ge 8$ with the 
ansatz~(\ref{simpleD}) we get $\Delta_b(10,8,1.07) = -0.000004(10)$
and $ -0.000012(10)$ for fixing $Z_a/Z_p=0.32037$ and $\xi_{2nd}/L=0.59238$, 
respectively. Fitting with the ansatz~(\ref{doubleD}), the estimates stay
compatible with zero, but with a larger error bar.  Taking also these
results into account we conclude that 
$| D^*(10) - D^*(8) |  \lessapprox 0.0005$. 

Next we have analyzed our data for $N=6$ and $D=1.02$.   Taking into account
the results of the fits using different ans\"atze, we arrive at
$\Delta_b(6,8,1.02) =0.00163(6)$ for fixing $Z_a/Z_p=0.32037$ and
$\Delta_b(6,8,1.02) =0.00198(13)$ for fixing $\xi_{2nd}/L=0.59238$.
Plugging in the numbers into eq.~(\ref{shiftD}) we arrive at
$D^*(6) - D^*(8) = 0.0132(5)$ and $0.0140(10)$, for fixing 
$Z_a/Z_p=0.32037$ or $\xi_{2nd}/L=0.59238$, respectively. As our final 
result we take 
\begin{equation}
D^*(6) - D^*(8) = 0.0136(14)
\end{equation}
covering both the results for fixing $Z_a/Z_p=0.32037$ and 
for fixing $\xi_{2nd}/L=0.59238$.

Assuming that $D^*(N)$ converges rapidly to $D^*(\infty)$,  we 
conclude that $| D^*(N) - D^*(8) |$ for $N>8$ is much smaller than the 
error of $D^*(8)$, eq.~(\ref{DSTAR}). 
It seems plausible that $D^*(7) - D^*(8)$ is smaller
than $D^*(6) - D^*(8)$ computed above.  Likely $|D^*(5) - D^*(8)|$
is considerably larger than the error of $D^*(8)$ and an effort beyond that of 
this section is required to obtained an accurate estimate of $D^*(5)$.

\subsection{$N$-dependence of the magnetic susceptibility and the slope of
dimensionless quantities}
\label{allonN}
Finally we have studied the dependence of quantities that we used to compute
the critical exponents $\nu$ and $\eta$ on $N$. In particular we consider 
the magnetic susceptibility and the slopes of dimensionless quantities 
at either $Z_a/Z_p=0.32037$ or $\xi_{2nd}/L=0.59238$.  Let us discuss the
results obtained for the susceptibility. Those for the 
slopes of dimensionless quantities are qualitatively the same.

We computed the ratio
\begin{equation}
\label{ratiochiN}
R_{\chi}(6,8) = \frac{\chi(N=6)}{\chi(N=8)}
\end{equation} 
for either  $Z_a/Z_p=0.32037$ or $\xi_{2nd}/L=0.59238$ fixed at $D=1.02$.
Following the discussion of section \ref{FSStheory}, this ratio should 
behave as
\begin{equation}
R_{\chi}(6,8) = a \; (1 + b L^{-\omega} + ...) \;,
\end{equation}
where all possible types of corrections should appear, 
and not only those related to the breaking of the $O(2)$ symmetry.
We have fitted our data by using a single correction term.
In the case of $Z_a/Z_p=0.32037$ we get the following results:

Using the correction exponent $\epsilon=0.789$ and $L_{min}=6$ we get
$a=1.000187(14)$, $b=0.00126(8)$ and $\chi^2/$d.o.f.$=0.36$. 
Using instead $\epsilon=2$ and $L_{min}=8$ we get
$a=1.000295(9)$,  $b=0.0091(10)$ and $\chi^2/$d.o.f.$=0.56$. 

For $\xi_{2nd}/L=0.59238$ fixed we get:
Using the correction exponent $\epsilon=0.789$ and $L_{min}=8$ we get
$a=1.000222(15)$, $b=-0.00010(11)$ and $\chi^2/$d.o.f.$=0.64$. 
Using instead $\epsilon=2$ and $L_{min}=4$ we get
$a=1.000219(4)$, $b=-0.00143(13)$ and $\chi^2/$d.o.f.$=0.59$.

We conclude that the ratio~(\ref{ratiochiN}) consists of an overall
constant that is close to one and corrections with a small amplitude.
Since these corrections come with a very small amplitude,
it is impossible to assign them clearly to the correction exponents
that are theoretically expected.

In the case of $N=8$ and $N=10$ at $D=1.07$ the data barely differ.
For example for $\chi$ at $\xi_{2nd}/L=0.59238$ we get for $L=4$ the
estimates $17.01708(4)$ and $17.01707(5)$, respectively. Therefore 
we abstain from any further analysis.

\section{The correction exponent $y_{N=6}$}
\label{appendixB}
We define
\begin{eqnarray}
 X_N &=& \langle \mbox{max}_j \vec{m} \vec{r}_j \rangle \;, \\
 Y_N &=& \langle \mbox{max}_j \vec{m} \vec{p}_j \rangle \;,
\end{eqnarray}
where 
\begin{eqnarray}
 \vec{r}_j &=&  \left(\cos(2 \pi j/N), \sin(2 \pi j/N) \right) \;, \\
 \vec{p}_j &=&  \left(\cos(2 \pi [j+1/2]/N), \sin(2 \pi [j+1/2]/N) \right) \;,
\end{eqnarray}
where $j \in \{0,...,N-1\}$ and 
\begin{equation}
(m^{(0)},m^{(1)}) = \vec{m} = \sum_x \vec{s}_x
\end{equation}
is the magnetization. 
Now we consider the quantity
\begin{equation}
\label{qnour}
 q_N = 
        \frac{X_N-Y_N}{X_N+Y_N} 
\end{equation}
as a measure of the deviation from $O(2)$ invariance. 
We performed simulations for  $N=6$ and $D=1.02$ close to our final estimate
of $D^*(6)=1.058(13)+0.0136(14)$. We simulated the lattice sizes
$L=4$, $5$, ..., $16$, $18$, $20$, and $64$, as discussed already  
above. 
The quantities $X_N$ and $Y_N$ are taken at $Z_a/Z_p=0.32037$. 
Note that $q_N$ for $L=64$ is equal to zero within 
error bars. Therefore we did not include $L=64$ in our analysis.
We fitted our numerical results with the ans\"atze
\begin{equation}
 q_N = c L^{y_{N=6}}
\end{equation}
and 
\begin{equation}
 q_N = c L^{y_{N=6}}  \times (1 + b L^{-2})  \;.
\end{equation}
We find $y_{N=6}=-2.42(2)$ and $\chi^2/$d.o.f.$=0.53$ with $L_{min}=8$ using 
the first ansatz and $y_{N=6}=-2.46(3)$ and $\chi^2/$d.o.f.$=0.59$ with $L_{min}=6$
using the second ansatz. As our final estimate we take $y_{N=6}=-2.43(6)$, where 
the error estimate includes the results of both fits.

This value has to be compared with $y_{N=6}=-2.55(6)$ and $-2.509(7)$
given in refs. \cite{Shao19} and \cite{Debasish}, respectively. 

Note that for $N>6$ it is virtually impossible to get a reliable 
estimate of $y_N$ using the method used here, since the relative 
error of $q_N$ is rapidly increasing with increasing $L$.

At a late stage of the project we have implemented the quantity 
\begin{equation}
 \phi_N = \langle \cos(N \Theta) \rangle \; ,
\end{equation}
where $\Theta=\arccos(m^{(0)}/|\vec{m}|)$, which is used
in ref. \cite{Shao19}; See eq.~(3) of ref. \cite{Shao19}. 
We simulated the linear lattice
sizes $L=4$, $6$, $8$, and $12$, measuring both $\phi_N$ and $q_N$. We find
that the relative error is slightly smaller for  $q_N$,
the two quantities are highly  correlated, and their ratio $\phi_N/q_N$ is 
within the statistical error the same for the lattice sizes $L=6$, $8$, and
$12$. For $L=4$ it deviates by little.
Hence for our purpose the two quantities are equivalent.

\def\refname{}


\begin{thebibliography}{99}
\bibitem{WiKo}
K. G. Wilson and J. Kogut,
{\sl The renormalization group and the $\epsilon$-expansion},
Phys.\ Rep.\ C {\bf 12}, 75 (1974).

\bibitem{Fisher74}
M. E. Fisher,
{\sl The renormalization group in the theory of critical behavior},
Rev.\ Mod.\ Phys.\ {\bf 46}, 597 (1974).

\bibitem{Fisher98}
M. E. Fisher,
{\sl Renormalization group theory: Its basis and formulation in statistical physics},
Rev.\ Mod.\ Phys.\ {\bf 70}, 653 (1998).

\bibitem{PeVi}
A. Pelissetto and E. Vicari,
{\sl Critical Phenomena and Renormalization-Group Theory},
[cond-mat/0012164],
Phys.\ Rept.\ {\bf 368}, 549 (2002).

\bibitem{Debasish}
D. Banerjee, S. Chandrasekharan, and D. Orlando,
{\sl Conformal dimensions via large charge expansion},
[arXiv:1707.00711], Phys.\ Rev.\ Lett.\ {\bf 120}, 061603 (2018).

\bibitem{HoSu03}
J. Hove and A. Sudb\o,
{\sl Criticality versus q in the (2+1)-dimensional $Z_q$ clock model},
[arXiv:cond-mat/0301499], 
Phys.\ Rev.\ E {\bf 68}, 046107 (2003).

\bibitem{Sandvik07}
Jie Lou, Anders W. Sandvik, and Leon Balents,
{\sl Emergence of U(1) Symmetry in the 3D XY Model with $Z_q$ Anisotropy},
[arXiv:0704.1472], Phys.\ Rev.\ Lett.\ {\bf 99}, 207203 (2007).

\bibitem{Lipa96}
J. A. Lipa, D. R. Swanson, J. A. Nissen, T. C. P. Chui, and U. E.
Israelsson,
{\sl  Heat Capacity and Thermal Relaxation of Bulk Helium very near the
Lambda Point}, Phys.\ Rev.\ Lett.\ {\bf 76}, 944  (1996).

\bibitem{Lipa00}
J. A. Lipa, D. R. Swanson, J. A. Nissen, Z. K. Geng, P. R. Williamson,
D. A. Stricker, T. C. P. Chui, U. E. Israelsson, and M. Larson,
{\sl Specific Heat of Helium Confined to a 57- $\mu$m Planar Geometry},
Phys.\ Rev.\ Lett.\  {\bf 84}, 4894 (2000).

\bibitem{Lipa03}
J. A. Lipa, J. A. Nissen, D. A. Stricker, D. R. Swanson and T. C. P. Chui,
{ \sl
Specific heat of liquid helium in zero gravity very near the $\lambda$-point},
[arXiv:cond-mat/0310163],
Phys.\ Rev.\ B {\bf 68}, 174518 (2003).

\bibitem{GuZi98}
R. Guida and J. Zinn-Justin,
{\sl Critical exponents of the N vector model},
[arXiv:cond-mat/9803240], J.\ Phys.\ A {\bf 31}, 8103 (1998).

\bibitem{XY2}
M. Campostrini, M. Hasenbusch, A. Pelissetto, and E. Vicari,
{\sl The critical exponents of the superfluid transition in He4},
[cond-mat/0605083], published as
{\sl Theoretical estimates of the critical exponents of the superfluid
transition in He4 by lattice methods}, Phys.\ Rev.\ B {\bf 74} (2006) 144506.

\bibitem{Xu19}
W. Xu, Y. Sun, J.-P. Lv, and Y. Deng, 
{\sl High-precision Monte Carlo study of several models in the 
three-dimensional U(1) universality class},
[arXiv:1908.10990], Phys.\ Rev.\ B {\bf 100}, 064525 (2019).

\bibitem{Kos:2016ysd}
F. Kos, D. Poland, D. Simmons-Duffin, and A. Vichi,
{Precision Islands in the Ising and $O(N)$ Models}
[arXiv:1603.04436], JHEP 08 (2016) 036.

\bibitem{che19}
S. M. Chester, W. Landry, J. Liu, D. Poland, D. Simmons-Duffin, N. Su, and A. Vichi,
{\sl Carving out OPE space and precise $O(2)$ model critical exponents},
[arXiv:1912.03324].

\bibitem{XY1}
M. Campostrini, M. Hasenbusch, A. Pelissetto, P. Rossi, and
E. Vicari,
{\sl Critical behavior of the three-dimensional XY universality class},
[cond-mat/0010360], Phys. Rev. B {\bf 63} (2001) 214503.

\bibitem{Simmons-Duffin:2016wlq}
 D. Simmons-Duffin,
{\sl The Lightcone Bootstrap and the Spectrum of the 3d Ising CFT},
[arXiv:1612.08471], JHEP 03 (2017) 086.

\bibitem{ChFiNi}
J. H. Chen, M. E. Fisher and B. G. Nickel,
{\sl Unbiased Estimation of Corrections to Scaling by Partial Differential
Approximants},
Phys.\ Rev.\ Lett.\ {\bf 48},
630 (1982).

\bibitem{FiCh}
M. E. Fisher and J. H. Chen,
{\sl The validity of hyperscaling in three dimensions for scalar spin systems},
 J.\ Physique (Paris) {\bf 46}, 1645 (1985).

\bibitem{Bloete}
 H. W. J. Bl\"ote, E. Luijten and J. R. Heringa,
{\sl Ising universality in three dimensions: a Monte Carlo study},
[arXiv:cond-mat/9509016],
J. Phys. A: Math. Gen. {\bf 28}, 6289 (1995).

\bibitem{Ballesteros}
H. G. Ballesteros, L. A. Fern\'andez, V. Mart\'in-Mayor, and A. Mu\~noz Sudupe,
{\sl Finite Size Scaling and “perfect” actions: 
the three dimensional Ising model},
[arXiv:hep-lat/9805022], Phys.\ Lett.\ B {\bf 441}, 330 (1998). 

\bibitem{KlausStefano}
M. Hasenbusch, K. Pinn, and S. Vinti,
{\sl Critical Exponents of the 3D Ising Universality Class From Finite Size Scaling With Standard and Improved Actions},
[arXiv:hep-lat/9806012], Phys.\ Rev.\ B {\bf 59}, 11471 (1999). 

\bibitem{Barber}
M. N. Barber, {Finite-size Scaling}
in {\sl Phase Transitions and Critical Phenomena, Vol. 8},
eds. C. Domb and J. L. Lebowitz, (Academic Press, 1983)

\bibitem{myPhi4}
M. Hasenbusch,
{\sl A Monte Carlo study of leading order scaling corrections 
of $\phi^4$ theory on a three dimensional lattice},
[hep-lat/9902026], J.\ Phys.\ A {\bf 32}, 4851 (1999). 

\bibitem{Tibor}
M. Hasenbusch and T. T\"or\"ok,
{\sl High precision Monte Carlo study of the 3D XY-universality class}
[arXiv:cond-mat/9904408], J.\ Phys.\ A {\bf 32}, 6361 (1999).

\bibitem{ourdilute}
M. Hasenbusch, F. Parisen Toldin, A. Pelissetto, and E. Vicari,
{\sl Universality class of 3D site-diluted and bond-diluted Ising systems},
[arXiv:cond-mat/0611707],
J.\ Stat. \ Mech.:\ Theory Exp. {\bf 2007}, P02016.

\bibitem{Ha10}
M. Hasenbusch,
{\sl A Finite Size Scaling Study of Lattice Models in the 3D
Ising Universality Class},
[arXiv:1004.4486],
Phys.\ Rev.\ B {\bf 82}, 174433 (2010).

\bibitem{Bi81}
K. Binder, 
{\sl Finite Size Scaling Analysis of
 Ising Model Block Distribution Functions}, 
Z.\ Phys.\ B: Condens. Matter {\bf 43}, 119 (1981).

\bibitem{myStructure}
M. Hasenbusch,
{\sl Two- and three-point functions at criticality:
Monte Carlo simulations of the improved three-dimensional Blume-Capel model},
[arXiv:1711.10946], Phys.\ Rev.\ E {\bf 97} (2018) 012119.

\bibitem{MaKrDi04}
A. Maciolek, M. Krech, and S. Dietrich,
{\sl Phase diagram of a model for $^3$He-$^4$He mixtures in three dimensions}, 
Phys.\ Rev.\ E {\bf 69}, 036117 (2004).

\bibitem{ROT98}
M. Campostrini, A. Pelissetto, P. Rossi, and E. Vicari,
{\sl Two-point correlation function of three-dimensional O(N) models: 
The critical limit and anisotropy}, [arXiv:cond-mat/9705086],
Phys. Rev. E {\bf 57}, 184 (1998).


\bibitem{Wolff}
U. Wolff,
{\sl Collective Monte Carlo Updating for Spin Systems},
Phys.\ Rev.\ Lett.\ {\bf 62}, 361  (1989).

\bibitem{Campostrini:2002cf}
 M. Campostrini,
 A. Pelissetto, P. Rossi,
 and E. Vicari,
{\sl 25th order high temperature expansion results for
   three-dimensional Ising like systems on the simple cubic
    lattice},
  [arXiv:cond-mat/0201180],
  Phys.\ Rev.\ E  {\bf 65}, 066127 (2002).

\bibitem{NewmanRiedel}
K. E. Newman and E. K. Riedel,
{\sl Critical exponents by the scaling-field method: The isotropic $N$-vector model in three dimensions},
Phys.\ Rev.\ B {\bf 30}, 6615 (1984).

\bibitem{Litim04}
D. F. Litim and L. Vergara,
{\sl Subleading critical exponents from the renormalisation group},
[arXiv:hep-th/0310101], Phys.\ Lett.\ B {\bf 581}, 263 (2004).

\bibitem{Litim17}
A. J\"uttner, D.F. Litim, and E. Marchais,
{\sl Global Wilson–Fisher fixed points},
[arXiv:1701.05168],
Nucl.\ Phys.\ B {\bf 921}, 769 (2017).




\bibitem{Weigel}
M. Weigel,
{\sl Simulating spin models on GPU}, 
Comput. Phys. Commun. {\bf 182}, 1833 (2011).

\bibitem{Komura}
Y. Komura and Y. Okabe, {\sl CUDA programs for GPU computing of Swendsen-Wang multi-cluster spin flip algorithm: 
2D and 3D Ising, Potts, and XY models},
[arXiv:1403.7560], Comput. Phys. Commun. {\bf 185}, 1038 (2014); ibid.,
{\sl Improved CUDA programs for GPU computing of Swendsen-Wang multi-cluster spin flip algorithm:
2D and 3D Ising, Potts, and XY models}, {\bf 200}, 400 (2016).

\bibitem{twister}
M. Saito and M. Matsumoto,
``SIMD-oriented Fast Mersenne Twister:
a 128-bit Pseudorandom Number Generator'',
in
{\sl Monte Carlo and Quasi-Monte Carlo Methods 2006},
edited by A. Keller, S. Heinrich, H. Niederreiter, (Springer, 2008);
M. Saito, Masters thesis, Math. Dept., Graduate School of science,
Hiroshima University, 2007.
The source code of the program is provided at
\verb+http://www.math.sci.hiroshima-u.ac.jp/~m-mat/MT/SFMT/index.html+

\bibitem{plotting}
J. D. Hunter, {\sl "Matplotlib: A 2D Graphics Environment}, 
Computing in Science \& Engineering {\bf 9}, 90 (2007).

\bibitem{pythonSciPy}
T. E. Oliphant, {\sl Python for Scientific Computing}, Computing in Science \& Engineering {\bf 9}, 10 (2007);
E. Jones,E. Oliphant, P. Peterson, et al.,{\sl SciPy: Open Source Scientific Tools for Python}, 2001-, 
\verb+http://www.scipy.org/+, 
P. Virtanen, R. Gommers, T. E. Oliphant et al.,
{\sl SciPy 1.0--Fundamental Algorithms for Scientific Computing in Python},
[arXiv:1907.10121].


\bibitem{LM1}
K. Levenberg, {\sl A method for the solution of certain non-linear problems in least squares}, 
Quart.\ Appl.\ Math.\ {\bf 2}, 164 (1944).

\bibitem{LM2}
D. Marquardt, {\sl An Algorithm for Least-Squares Estimation of Nonlinear Parameters}, 
SIAM J.\ Appl.\ Math.\ {\bf 11}, 431, (1963).

\bibitem{LM3}
J. J. Mor\'e, {\sl The Levenberg-Marquardt algorithm: Implementation and theory}, in G. A. Watson (ed.): 
Numerical Analysis. Dundee 1977, Lecture Notes Math. {\bf 630}, 105 (1978).

\bibitem{MINPACK}
J. J. Mor\'e, B. S. Garbow, and K. E. Hillstrom, {\sl User Guide for MINPACK-1}, 
Argonne National Laboratory Report ANL-80-74, Argonne, Ill., (1980);
J. J. Mor\'e, D. C. Sorensen, K. E. Hillstrom, and B. S. Garbow, {\sl The MINPACK Project}, 
in Sources and Development of Mathematical Software, W. J. Cowell, ed., Prentice-Hall, 88 (1984).

\bibitem{Deng05}
Y.J. Deng, H.W.J. Bl\"ote, M.P. Nightingale, 
{\sl Surface and bulk transitions in three-dimensional O(n) models},
[arXiv:cond-mat/0504173], Phys. Rev. E 72, 016128 (2005).

\bibitem{Lan12}
T.-Y. Lan, Y.-D. Hsieh, and Y.-J. Kao,
{\sl  High-precision Monte Carlo study of the three-dimensionalXY model on GPU},
[arXiv:1211.0780].

\bibitem{Shao19}
H. Shao, W. Guo, and A. W. Sandvik, 
{\sl Monte Carlo Renormalization Flows in the Space of Relevant and Irrelevant Operators: 
Application to Three-Dimensional Clock Models},
[arXiv:1905.13640].


\end{thebibliography}
\end{document}